\documentclass[12pt,draftclsnofoot,onecolumn]{IEEEtran}
\IEEEoverridecommandlockouts
\usepackage{color}
\usepackage{graphicx}
\usepackage{epstopdf}
\usepackage{amsmath}
\usepackage{amssymb}
\usepackage{algorithm}
\usepackage{algorithmic}
\usepackage{amsmath}
\usepackage{multirow}
\usepackage{booktabs}
\usepackage{array}
\usepackage{amsthm}
\usepackage{stfloats}
\usepackage{caption}
\usepackage{subfigure}
\usepackage{bm}
\usepackage{booktabs}
\usepackage{setspace}
\usepackage{diagbox}
\usepackage{enumerate}
\usepackage{ulem}
\usepackage{url}
\usepackage{circledsteps}

\usepackage{hyperref}
\allowdisplaybreaks[4]

{\bgroup
 \addtolength\abovedisplayshortskip{#1}
 \addtolength\abovedisplayskip{#1}
 \addtolength\belowdisplayshortskip{#1}
 \addtolength\belowdisplayskip{#1}
 }
{\egroup\ignorespacesafterend}

\newtheorem{theorem}{Theorem}
\newtheorem{lemma}{Lemma}

\newcommand{\be}{\begin{equation}}
\newcommand{\ee}{\end{equation}}
\newcommand{\bea}{\begin{eqnarray}}
\newcommand{\eea}{\end{eqnarray}}
\newcommand{\ba}{\begin{array}}
\newcommand{\ea}{\end{array}}

\title{
Transmissive RIS Transceiver-Empowered ISAC Systems: Energy Efficiency Optimization for Perfect and Imperfect CSI
}
\author{\IEEEauthorblockN{Yuan Guo, Wen Chen, Yang Liu, Kunlun Wang, Zhendong Li, and Qiong Wu
}
\thanks{
Y. Guo and W. Chen are with the Department of Electronic Engineering, Shanghai Jiao Tong University, Shanghai 200240, China 
(e-mail:
yuanguo26@sjtu.edu.cn;
wenchen@sjtu.edu.cn).}
\thanks{
Y. Liu  is with the School of Information
and Communication Engineering, Dalian University of Technology, Dalian 116024, China 
(e-mail:
yangliu\_613@dlut.edu.cn).}
\thanks{
K. Wang is with 
the School of Communication and Electronic Engineering, 
East China Normal University, Shanghai 200000, China
(e-mail: klwang@cee.ecnu.edu.cn).
}
\thanks{Z. Li is with the School of Information and Communication Engineering, Xi'an Jiaotong University, Xi'an 710049, China
(e-mail:
lizhendong@xjtu.edu.cn).}
\thanks{
Q. Wu is with the School of Internet of Things Engineering, 
Jiangnan University, Wuxi 214122, China 
(e-mail: qiongwu@jiangnan.edu.cn).
}
}

\begin{document}
\maketitle
\pagestyle{empty}
\thispagestyle{empty}

\begin{abstract}
In this paper,
a novel transmissive reconfigurable intelligent surface (TRIS) transceiver is employed to enable an integrated sensing and communication (ISAC) system supporting both communication and sensing.
Under both perfect and imperfect channel state information (CSI), 
we study the transmit beamforming design for the TRIS transceiver to maximize the system energy efficiency (EE), 
subject to per-user minimum-rate guarantees, 
a minimum beampattern gain toward the sensing target, 
and per-antenna power constraints.
The corresponding EE maximization problems are challenging to solve due to the fractional objective and non-convex constraints. 
In particular, 
under imperfect CSI, 
the resulting semi-infinite constraints further complicate the problem.
For the perfect CSI case, 
we first apply the fractional programming (FP) methodology to obtain more tractable reformulations of the rate functions, 
and then propose an iterative algorithm based on the majorization-minimization (MM) framework. 
For the imperfect CSI case, 
we utilize the S-Procedure to transform the semi-infinite inequality constraints into linear matrix inequalities (LMIs), 
and further develop an efficient MM-based algorithm with the aid of slack variables. 
Numerical results demonstrate the convergence and effectiveness of the proposed algorithms 
and validate the EE gains of the TRIS transceiver-enabled ISAC system.

\end{abstract}


\maketitle
\section{Introduction}

The forthcoming sixth-generation (6G) wireless networks, 
featuring ubiquitous coverage, low latency, high data rates, and integrated sensing capability, 
are expected to enable a broad range of emerging services, 
such as vehicle-to-everything (V2X) communications, low-altitude applications, and the Internet of Things (IoT) \cite{ref_ISAC_1}.
In this context, 
integrated sensing and communication (ISAC), 
as one of the critical technologies for 6G, 
has attracted considerable interest from both academia and industry \cite{ref_ISAC_2}.
In particular, 
ISAC  allows communication and sensing to share the same radio resources (i.e., time, frequency, and spatial domains) 
and hardware platform,
which can improve spectral and energy efficiencies while reducing hardware cost.
Moreover, by jointly designing the communication and sensing functionalities, 
the two functionalities can complement each other instead of being optimized independently.
Therefore, ISAC is considered a potential solution for 6G by combining the two capabilities and exploiting their synergy.
Accordingly, 
recent works on ISAC design can be found in \cite{ref_ISAC_2}$-$\cite{ref_ISAC_5} and the references therein.

Recently, 
reconfiguring the wireless propagation environment has emerged as a pivotal research direction in wireless communications.
As a representative paradigm, 
the reconfigurable intelligent surface (RIS) has attracted significant attention due to its capability of programmably manipulating electromagnetic (EM) waves \cite{ref_RIS_1}.
In general,
an RIS consists of a two-dimensional metasurface equipped with a large number of low-cost and passive tunable units,
each of which can be controlled to adjust the phase and/or amplitude of the impinging EM waves.
Due to its unique functionality,
the RIS can shape the wireless channel to enhance the desired signal and mitigate interference, 
thereby improving coverage and spectral efficiency.
Thus, 
the RIS is regarded as a promising solution for realizing energy-efficient and cost-effective wireless networks.
Extensive studies on RIS-assisted networks can be found in \cite{ref_RIS_1}$-$\cite{ref_RIS_ISAC_1} and the references therein.
Specifically, 
related works on RIS-assisted ISAC systems include \cite{ref_RIS_ISAC_0}$-$\cite{ref_RIS_ISAC_1} and the references therein.

However, most of the existing literature \cite{ref_RIS_1}$-$\cite{ref_RIS_ISAC_1} has focused on conventional RIS-assisted wireless networks, 
in which the RIS mainly assists external transceivers by passively reflecting and reconfiguring impinging electromagnetic waves. 
By contrast, 
a novel transmissive RIS (TRIS)-based transceiver architecture, 
referred to as the TRIS transceiver, 
was introduced in \cite{ref_TRIS_1}. 
This architecture mainly consists of a passive transmissive RIS and a single horn-feed antenna.
Specifically,
the feed antenna illuminates the transmissive surface with an incident carrier wave. 
Then, the controller modulates the state of each unit based on time-modulated array (TMA) technology, 
thereby shaping the transmitted wave.
Moreover, 
compared with conventional multi-antenna transceivers, 
the TRIS transceiver can reduce hardware complexity by avoiding the use of a large number of RF chains and complicated signal processing modules. 
Furthermore, 
since the horn-feed antenna and the served receivers are located on opposite sides of the TRIS, 
the incident and transmitted waves are naturally separated in space. 
Therefore, 
compared with reflective-type RIS transmitter architectures \cite{ref_RIS_transceiver_1}, 
the TRIS transceiver can effectively mitigate both feed blockage and echo interference.

Motivated by its unique radiation and coverage characteristics,
the TRIS transceiver has attracted increasing interest in a variety of emerging applications
\cite{ref_TRIS_app_1}$-$\cite{ref_TRIS_app_10}.
For instance,
the authors of \cite{ref_TRIS_app_1} investigated a TRIS transceiver-enabled multi-stream communication system and proposed a linear-complexity beamforming algorithm to solve a max-min signal-to-interference-plus-noise ratio (SINR) optimization problem, 
thereby demonstrating the benefits of the TRIS transceiver architecture.
The papers \cite{ref_TRIS_app_1_1}$-$\cite{ref_TRIS_app_1_2} investigated a TRIS transceiver-assisted simultaneous wireless information and power transfer (SWIPT) system.
Based on plane-wave and spherical-wave propagation models and considering imperfect channel state information (CSI),
the paper \cite{ref_TRIS_app_2} studied a robust sum-rate maximization problem in a TRIS transceiver-empowered SWIPT network.
To address resource constraints in multi-tier computing networks, including limited computing capability, stringent latency requirements, and power constraints, 
the work \cite{ref_TRIS_app_3} adopted a TRIS transceiver to mitigate these limitations.
The literature \cite{ref_TRIS_app_4} firstly proposed a novel hybrid active-passive TRIS transmitter, 
where each unit can dynamically switch between on/off states. 
Furthermore, an energy efficiency (EE) maximization problem was formulated to evaluate the effectiveness of the proposed architecture.
To improve service coverage in ISAC systems, 
the authors in \cite{ref_TRIS_app_5} jointly adopted a TRIS transceiver and rate-splitting multiple access (RSMA) technique, 
and investigated an optimization problem that maximizes the minimum radar mutual information (RMI), 
in order to quantify the cooperative gains.
Since power consumption is a critical issue in integrated sensing, computing, and communication (ISCC) networks, 
the work \cite{ref_TRIS_app_6} employed a TRIS transceiver to reduce the system's total power consumption.
In \cite{ref_TRIS_app_7} and \cite{ref_TRIS_app_8}, 
the TRIS transceiver was exploited to improve transmission performance in multicast and multicell communication systems, respectively.
To address stringent power constraints in cognitive multi-input multi-output (MIMO) radar systems,
the paper \cite{ref_TRIS_app_9} leveraged a TRIS transceiver to reduce power consumption.
In \cite{ref_TRIS_app_10},
an innovative base-station (BS) architecture integrating movable antennas (MAs) and a TRIS transceiver was proposed
to improve the signal-to-noise ratio (SNR).

\subsection{Motivations and Contributions}

First,
future 6G wireless networks are expected to provide high data rates and integrated sensing capabilities while supporting sustainable low-power operation.
Therefore,
EE has become a critical performance metric.
However, 
most existing studies on wireless networks enabled by the TRIS transceiver mainly focus on conventional performance metrics,
such as 
SNR \cite{ref_TRIS_app_10},
SINR \cite{ref_TRIS_app_1}, 
sum-rate \cite{ref_TRIS_app_1_1}$-$\cite{ref_TRIS_app_2}, \cite{ref_TRIS_app_7}$-$\cite{ref_TRIS_app_8}, 
total power consumption \cite{ref_TRIS_app_3}, \cite{ref_TRIS_app_6}, 
RMI \cite{ref_TRIS_app_5}, 
or beampattern gain \cite{ref_TRIS_app_9}.
Consequently, 
EE-oriented design for TRIS-enabled wireless networks remains largely underexplored.
Moreover, 
most of the existing literature 
\cite{ref_TRIS_app_1}$-$\cite{ref_TRIS_app_1_1}, \cite{ref_TRIS_app_3}, \cite{ref_TRIS_app_5}$-$\cite{ref_TRIS_app_10}
assumes perfect CSI, 
whereas CSI is usually imperfect in practical systems due to estimation errors and/or feedback quantization \cite{ref_channel_error_1}. 
Note that \cite{ref_TRIS_app_4} considered robust EE design only for a TRIS transceiver-enabled communication system. 
As discussed above, 
the robust EE optimization for TRIS transceiver-empowered ISAC systems remains underexplored and is therefore desirable.
Specifically,
the key contributions of this paper are summarized as follows:
\begin{itemize}
\item
This paper investigates the transmit beamforming design for an ISAC system enabled by the novel TRIS transceiver 
to improve EE while guaranteeing both communication and sensing performance.
Specifically,
the goal of this paper is to maximize the EE of the considered ISAC system,
subject to minimum rate requirements for all mobile users,
a predefined beampattern gain threshold at the sensing target,
and the per-antenna power constraints of the TRIS transceiver.
Moreover, we consider this EE maximization problem for two CSI cases, 
i.e.,
perfect CSI and imperfect CSI.
To the best of our knowledge,  
this problem has not yet been considered in the existing literature, 
e.g., \cite{ref_TRIS_app_1}$-$\cite{ref_TRIS_app_10}.

\item
For the perfect CSI case, 
we develop a tailored algorithm for the considered EE maximization problem, 
which is challenging due to the fractional objective and the non-convex constraints. 
Specifically, 
we first equivalently transform the rate expressions in both the objective and the constraints into a more tractable form by invoking the fractional programming (FP) framework \cite{ref_FP}.
Then, 
by introducing slack variables and leveraging the majorization-minimization (MM) method \cite{ref_MM}, 
an efficient iterative algorithm is obtained that guarantees a feasible solution and monotonic convergence of the objective value.

\item
Furthermore,
under imperfect CSI,
the channel uncertainty induced by the bounded channel error model significantly increases the difficulty of the EE maximization problem compared with the perfect CSI case.
Specifically,
to address the semi-infinite uncertainty constraints,
we utilize the S-Procedure \cite{ref_S_Procedure} to convert these semi-infinite constraints into a finite set of linear matrix inequalities (LMIs).
Then, 
by introducing slack variables and leveraging the MM method,
we propose a tailored solution.

\item
Last but not least,
extensive simulation results are provided to validate the effectiveness and efficiency
of our proposed algorithms under various system parameters.
Moreover, 
the TRIS transceiver is shown to bring improved overall system performance.

\end{itemize}

\section{System Model and Problem Formulation}
\subsection{System Model}

\begin{figure}[t]
	\centering
	\includegraphics[width=.40\textwidth]{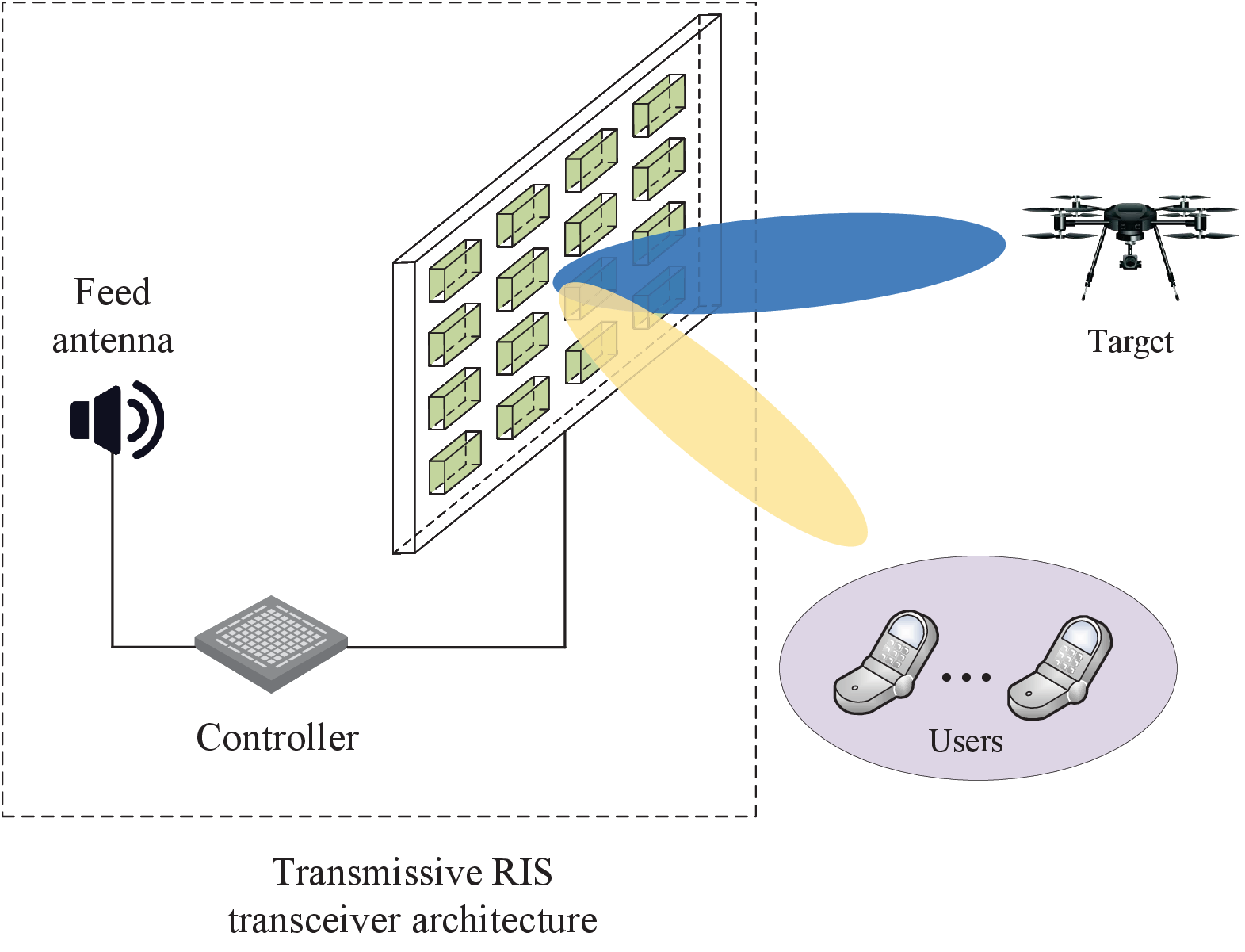}
	\caption{A TRIS transceiver-enabled ISAC system.}
	\label{fig.1}
\end{figure}

As shown in Fig. \ref{fig.1},
we consider a TRIS transceiver-enabled downlink ISAC system,
which consists of one TRIS transceiver, 
$K$ single-antenna mobile users, and one point-like sensing target.
It is assumed that the TRIS transceiver is equipped with $N$ units.
In the considered TRIS-enabled ISAC system, 
the TRIS transceiver achieves simultaneous dual-function transmission. 
Specifically, 
it transmits probing and communication waveforms for target sensing and information delivery, respectively.
To simplify the notation, 
we collect the users and the units of the TRIS transceiver in sets 
$\mathcal{K} = \{1,2,\cdots,K\} $ and $\mathcal{N} = \{1,2,\cdots,N\}$, respectively.

The signal transmitted by the TRIS transceiver
\footnote{
{This signal model can fully exploit the available degrees of freedom (DoFs) \cite{ref_joint_signal}.}
} is given as
\begin{align}
\mathbf{x} = {\sum}_{k=1}^{K} \mathbf{w}_{c,k}x_{c,k}
+
{\sum}_{n=1}^{N} \mathbf{w}_{r,n}x_{r,n},
\end{align}
where
$x_{c,k} \in \mathbb{C}$
and 
$x_{r,n} \in \mathbb{C}$ 
are communication-oriented information-bearing
and sensing-oriented probing signals,
respectively.
Without loss of generality,
we assume that $\mathbb{E}\{ \vert x_{c,k}  \vert^2 \} = 1$, 
$\mathbb{E}\{ \vert x_{r,n}  \vert^2 \} = 1$,
and 
$\mathbb{E}\{   x_{c,k}x_{r,n}^H   \} = 0$,
$\forall k \in \mathcal{K}, \forall n \in \mathcal{N}$.
$\mathbf{w}_{c,k}\in \mathbb{C}^{N\times 1}$
and
$\mathbf{w}_{r,n}\in \mathbb{C}^{N\times 1}$
denote the beamforming vectors for serving the $k$-th user and for target sensing, respectively.

Following the operating principle of the TRIS transceiver \cite{ref_TRIS_1},
the beamforming vectors are required to satisfy the following per-antenna power constraints,
which can be given by
\begin{align}
{\sum}_{k=1}^{K}
\mathbf{w}_{c,k}^H {\mathbf{A}}_{n} \mathbf{w}_{c,k} 
+ {\sum}_{i=1}^{N}\mathbf{w}_{r,i}^H {\mathbf{A}}_{n}\mathbf{w}_{r,i} 
\leq P_{unit}, \forall n \in \mathcal{N},
\end{align}
where $P_{unit}$ 
denotes the maximum transmit power available at the TRIS transceiver unit.
The selection matrix $\mathbf{A}_n$ can be formulated as
\begin{align}
\mathbf{A}_n \triangleq \text{diag}(\mathbf{a}_{n}) \in \mathbb{R}^{N\times N},
\end{align}
and the index vector $\mathbf{a}_{n}$ is given as
\begin{align}
\mathbf{a}_{n} \triangleq [0,0,\cdots,\underbrace{1}\limits_{\textrm{n-}th},\cdots,0,0]^T \in \mathbb{R}^{N\times 1},
\end{align}
where the element at position 
$n$ equals $1$ and all other elements are $0$.

\textit{1) Communication Model:}
In this work, 
all channels associated with the TRIS transceiver are modeled as quasi-static block-fading channels.
Focusing on a specific fading block, 
we assume that the channels remain constant over the block duration.
The channel from the TRIS transceiver to the $k$-th user is denoted by
$\mathbf{h}_{c,k} \in \mathbb{C}^{N \times 1}$ \footnote{The channel can be estimated using the methods proposed in \cite{ref_channel_estimation_1}$-$\cite{ref_channel_estimation_2}. }.
The number of TRIS units is given by
$
N = N_h N_v,
$
where \(N_h\) and \(N_v\) denote the numbers of TRIS units in the horizontal and vertical directions, respectively.
The channel from the TRIS to user \(k\) is modeled as a Rician fading channel, which is expressed as
\begin{align}
\mathbf{h}_{c,k}
=
\sqrt{C_0 \left(\frac{d_{c,k}}{d_0}\right)^{-\alpha}}
\left(
\sqrt{\frac{\kappa}{\kappa+1}}\,\mathbf{h}_{c,k}^{\mathrm{LoS}}
+
\sqrt{\frac{1}{\kappa+1}}\,\mathbf{h}_{c,k}^{\mathrm{NLoS}}
\right),
\end{align}
where \(C_0\) is the large-scale fading coefficient at the reference distance \(d_0=1\,\mathrm{m}\), 
\(d_{c,k}\) is the distance between the TRIS and user \(k\), 
\(\alpha\) is the path-loss exponent, 
and \(\kappa\) is the Rician factor.
The term \(\mathbf{h}_{c,k}^{\mathrm{LoS}}\) denotes the line-of-sight (LoS) component of the channel between the TRIS transceiver and the \(k\)-th user, 
which is given as (\ref{h_LoS}),
\begin{figure*}
\begin{small}
\begin{align}
\mathbf{h}_{c,k}^{\mathrm{LoS}}
&=
\bigg[
1,
e^{-j \frac{2\pi}{\lambda} d_s \sin \theta_{c,k}^{\mathrm{AoD}} \cos\psi_{c,k}^{\mathrm{AoD}}},
\cdots,
e^{-j \frac{2\pi}{\lambda} (N_h-1) d_s \sin \theta_{c,k}^{\mathrm{AoD}} \cos\psi_{c,k}^{\mathrm{AoD}}}
\bigg]^T \label{h_LoS} \\
&\quad \otimes
\bigg[
1,
e^{-j \frac{2\pi}{\lambda} d_s \cos \theta_{c,k}^{\mathrm{AoD}} },
\cdots,
e^{-j \frac{2\pi}{\lambda} (N_v-1) d_s \cos \theta_{c,k}^{\mathrm{AoD}} }
\bigg]^T, \nonumber
\end{align}
\end{small}
\boldsymbol{\hrule}
\end{figure*}
where \(\theta_{c,k}^{\mathrm{AoD}}\) and \(\psi_{c,k}^{\mathrm{AoD}}\) denote the vertical and horizontal angles of departure (AoDs) 
from the TRIS transceiver to user \(k\), respectively, 
\(d_s\) is the spacing between adjacent TRIS units, and \(\lambda\) is the carrier wavelength.
Moreover, the \(((n_h-1)N_v+n_v)\)-th entry of the NLoS component \(\mathbf{h}_{c,k}^{\mathrm{NLoS}}\) follows
\(
\mathbf{h}_{c,k}^{\mathrm{NLoS}}\big[(n_h-1)N_v+n_v\big] \sim \mathcal{CN}(0,1).
\)

The signal received at the $k$-th mobile user can be given as
\footnote{
{Different from the conventional RIS-aided system, 
whose signal model involves the cascaded transmitter-RIS-user channel together with the RIS reflection coefficients, 
the TRIS-enabled system only considers the direct TRIS-user channel.}
}
\begin{align}
{y}_{c,k}
=
\underbrace{{\mathbf{h}}_{c,k}^H\mathbf{w}_{c,k}x_{c,k}}
\limits_{\textrm{Desired signal}}
+\underbrace{{\sum}_{i\neq k}^{K}{\mathbf{h}}_{c,k}^H\mathbf{w}_{c,i}x_{c,i}}
\limits_{\textrm{Other users' interference}} 
+ \underbrace{{\sum}_{n=1}^{N} {\mathbf{h}}_{c,k}^H\mathbf{w}_{r,n}x_{r,n}}
\limits_{\textrm{Radar sensing interference}} 
 + n_{c,k},
\end{align}
where
$n_{c,k} \sim \mathcal{CN}(0, \sigma_{c,k}^2)  $ denotes the complex additive white Gaussian noise (AWGN) at the $k$-th user.

Thus, 
the SINR of user $k$ can be given as
\begin{align}
&\text{SINR}_{k}(\{\mathbf{w}_{c,k}\}, \{\mathbf{w}_{r,n}\})
=\frac{\vert{\mathbf{h}}_{c,k}^H\mathbf{w}_{c,k}  \vert^2}
{\sum_{i\neq k}^{K}\vert {\mathbf{h}}_{c,k}^H\mathbf{w}_{c,i}\vert^2
+\sum_{n=1}^{N} \vert {\mathbf{h}}_{c,k}^H\mathbf{w}_{r,n}\vert^2
+
\sigma_{c,k}^2}.
\end{align}

Accordingly, 
the corresponding achievable rate of the $k$-th user is defined as
\begin{align}
&\mathrm{R}_{k}(\{\mathbf{w}_{c,k}\}, \{\mathbf{w}_{r,n}\})
 =\text{log}\big( 1 + \text{SINR}_{k} (\{\mathbf{w}_{c,k}\}, \{\mathbf{w}_{r,n}\})  \big),
\forall k \in \mathcal{K}.
\end{align}

\textit{2) Radar Sensing Model:}
First,
the channel between the TRIS transceiver and the target is modeled as the LoS channel model,
which can be formulated in (\ref{h_r_LoS}),
\begin{figure*}
\begin{align}
{\mathbf{h}_{r}}
=&\alpha_{r}
\bigg[
1,
e^{-j \frac{2\pi}{\lambda} d_s \sin \theta_{r}^{\mathrm{AoD}} \cos\psi_{r}^{\mathrm{AoD}}  },
\cdots,
e^{-j \frac{2\pi}{\lambda} (N_h-1) d_s \sin \theta_{r}^{\mathrm{AoD}} \cos\psi_{r}^{\mathrm{AoD}}  }
\bigg]^T \label{h_r_LoS} \\
&\otimes\bigg[
1,
e^{-j \frac{2\pi}{\lambda} d_s \cos \theta_{r}^{\mathrm{AoD}}   },
\cdots,
e^{-j \frac{2\pi}{\lambda} (N_v-1) d_s \cos \theta_{r}^{\mathrm{AoD}}   }
\bigg]^T,\nonumber
\end{align}
\boldsymbol{\hrule}
\end{figure*}
where
$\theta_{r}^{\mathrm{AoD}}$
and
$\psi_{r}^{\mathrm{AoD}}$
denote the vertical and horizontal AoDs from the TRIS transceiver to the target, respectively,
and $\alpha_{r}$ represents the complex channel coefficient of the sensing link.

{
Next,
the incident signal impinging on the target can be formulated as}
\begin{align}
{{s}_{r}
=
{\sum}_{i=1}^{N} {\mathbf{h}}_{r}^H\mathbf{w}_{r,i}x_{r,i}
+
{\sum}_{k=1}^{K}{\mathbf{h}}_{r}^H\mathbf{w}_{c,k}x_{c,k}.}
\end{align}

Thus,
the beampattern gain \cite{ref_beampattern}$-$\cite{ref_beampattern_1} directed at the target is given by
\begin{align}
\mathrm{P}_{target}(\{\mathbf{w}_{c,k}\}, \{\mathbf{w}_{r,n}\})
&=\mathbb{E}\bigg\{
\bigg\vert 
{\sum}_{i=1}^{N}{\mathbf{h}}_{r}^H   \mathbf{w}_{r,i}x_{r,i}
+{\sum}_{k=1}^{K}{\mathbf{h}}_{r}^H \mathbf{w}_{c,k}x_{c,k}
\bigg\vert^2
\bigg\}\\
&= {\sum}_{i=1}^{N}\vert {\mathbf{h}}_{r}^H \mathbf{w}_{r,i} \vert^2
+{\sum}_{k=1}^{K}\vert {\mathbf{h}}_{r}^H \mathbf{w}_{c,k}\vert^2.\nonumber
\end{align}

{
\textit{3) Power Consumption Model:}
The power consumption of the TRIS transceiver is given as
\begin{align}
&\mathrm{P}_{TRIS}(\{\mathbf{w}_{c,k}\},\{\mathbf{w}_{r,n}\}) \\
&=\xi_{TRIS}{\sum}_{n=1}^{N}\bigg( {\sum}_{k=1}^{K}\mathbf{w}_{c,k}^H {\mathbf{A}}_{n} \mathbf{w}_{c,k}
+ {\sum}_{i=1}^{N}\mathbf{w}_{r,i}^H {\mathbf{A}}_{n} \mathbf{w}_{r,i} \bigg) +P_{c,TRIS},\nonumber
\end{align}
where 
$\xi_{TRIS}$ denotes the inverse of 
the transmit power amplifier efficiency of the TRIS unit circuitry.
The circuit power consumption of the TRIS transceiver is defined as
$P_{c,{TRIS}} \triangleq P_{s,{TRIS}}+P_{{RF}}$,
where $P_{s,{TRIS}}$ denotes the static circuit power consumption of the TRIS transceiver, 
and
$P_{{RF}}$ represents the power consumption of the single RF chain \cite{ref_P_cir}.}

\textit{4) Bounded CSI Error Model:}
First,
the channels 
$\mathbf{h}_{c,k}$
and 
$\mathbf{h}_{r}$
can be represented as
\begin{align}
&\mathbf{h}_{c,k} = \hat{\mathbf{h}}_{c,k} + \Delta\mathbf{h}_{c,k},
\mathbf{h}_{r} = \hat{\mathbf{h}}_{r} + \Delta\mathbf{h}_{r},
\end{align}
respectively,
{
where
$\hat{\mathbf{h}}_{c,k}$ and $\hat{\mathbf{h}}_{r}$ are the estimated channel vectors known at the TRIS transceiver,
$\Delta\mathbf{h}_{c,k}$ denotes the unknown communication channel estimation error vector, 
and $\Delta\mathbf{h}_{r}$ is the unknown sensing channel uncertainty vector induced by target angular errors.}
Following the worst-case robust design paradigm, 
we model the CSI errors using the bounded uncertainty sets \cite{ref_channel_error_model}$-$\cite{ref_channel_error_model_1}, i.e.,
\begin{align}
\Vert\Delta\mathbf{h}_{c,k}\Vert_2 \leq \xi_{c,k},
\Vert\Delta\mathbf{h}_{r} \Vert_2 \leq \xi_{r},
\end{align}
where
$\xi_{c,k}$
and
$\xi_{r}$ \footnote{ {The radius $\xi_r$ can be conservatively determined as an upper bound on the channel deviation over the angular uncertainty region \cite{ref_channel_error_model_2}.}} 
denote the radii of the uncertainty regions known at the TRIS transceiver.
Let
$\mathcal{H}_{c,k} = \{\mathbf{h}_{c,k} \vert
\mathbf{h}_{c,k} = \hat{\mathbf{h}}_{c,k} + \Delta\mathbf{h}_{c,k},
\Vert\Delta\mathbf{h}_{c,k}\Vert_2 \leq \xi_{c,k} \}$
and
$\mathcal{H}_{r} = \{\mathbf{h}_{r} \vert
\mathbf{h}_{r} = \hat{\mathbf{h}}_{r} + \Delta\mathbf{h}_{r},
\Vert\Delta\mathbf{h}_{r}\Vert_2 \leq \xi_{r} \}$.

\subsection{Problem Formulation}
In this paper,
we aim to maximize the system EE,
subject to the constraints that the achievable rate of each user is no less than a pre-defined threshold, 
the beampattern gain of the target is above a given threshold, 
and the per-antenna transmit power at each TRIS transceiver unit does not exceed a pre-defined maximum power. 
Depending on whether the TRIS transceiver has perfect CSI of the communication channels \(\{\mathbf{h}_{c,k}\}\) and the sensing channel \(\mathbf{h}_r\), 
we consider the following two cases:

\textit{1) Perfect CSI:}
In this case,
the TRIS transceiver has the perfect CSI of all channels,
and the corresponding optimization problem is mathematically formulated as
\begin{subequations}
\begin{align}
&\textrm{(P0)}:\mathop{\textrm{max}}
\limits_{\{\mathbf{w}_{c,k}\},\{\mathbf{w}_{r,n}\}
}\
\frac{
{\sum}_{k=1}^{K}
\mathrm{R}_{k}(\{\mathbf{w}_{c,k}\},\{\mathbf{w}_{r,n}\}) }
{\mathrm{P}_{TRIS}(\{\mathbf{w}_{c,k}\},\{\mathbf{w}_{r,n}\})}
\label{P0_obj}\\
&\textrm{s.t.}\ 
 \mathrm{R}_{k}(\{\mathbf{w}_{c,k}\},\{\mathbf{w}_{r,n}\}) \geq R_{th}, \forall k \in \mathcal{K},\label{P0_c_1}\\
&  \mathrm{P}_{target}(\{\mathbf{w}_{c,k}\},\{\mathbf{w}_{r,n}\})  \geq P_r, \label{P0_c_2}\\
& {\sum}_{k=1}^{K}
\mathbf{w}_{c,k}^H {\mathbf{A}}_{n} \mathbf{w}_{c,k} 
+ {\sum}_{i=1}^{N}\mathbf{w}_{r,i}^H {\mathbf{A}}_{n}\mathbf{w}_{r,i} 
\leq P_{unit},  \label{P0_c_3} 
\end{align}
\end{subequations}
where $R_{th}$ denotes the minimum rate requirement of the $k$-th user,
$P_r$ is the minimum beampattern gain,
and $P_{unit}$ represents the maximum transmit power of each TRIS unit.

\textit{2) Imperfect CSI:}
In this case, 
the perfect CSI of all channels is not available at the TRIS transceiver,
and the corresponding worst-case problem is given as
\begin{subequations}
\begin{align}
&\textrm{(P1)}:\mathop{\textrm{max}}
\limits_{\{\mathbf{w}_{c,k}\},\{\mathbf{w}_{r,n}\}
}\
\frac{ {\sum}_{k=1}^{K}\min_{\mathbf{h}_{c,k}\in\mathcal{H}_{c,k}} \mathrm{R}_{k} } {\mathrm{P}_{TRIS}}
\label{P1_obj}\\
&\textrm{s.t.}\ 
 \mathrm{R}_{k}(\{\mathbf{w}_{c,k}\},\{\mathbf{w}_{r,n}\})\! \geq\! R_{th}, 
 \forall \mathbf{h}_{c,k} \in  \mathcal{H}_{c,k}, 
 \forall k \!\in\! \mathcal{K},\label{P1_c_1}\\
&  \mathrm{P}_{target}(\{\mathbf{w}_{c,k}\},\{\mathbf{w}_{r,n}\})  \geq P_r, 
 \forall \mathbf{h}_{r} \in  \mathcal{H}_{r}, \label{P1_c_2}\\
& {\sum}_{k=1}^{K}
\mathbf{w}_{c,k}^H {\mathbf{A}}_{n} \mathbf{w}_{c,k} 
+ {\sum}_{i=1}^{N}\mathbf{w}_{r,i}^H {\mathbf{A}}_{n}\mathbf{w}_{r,i} 
\leq P_{unit}.  \label{P1_c_3} 
\end{align}
\end{subequations}

{
It is worth noting that $\mathrm{P}_{TRIS}$ is independent of the channel realization, 
since it is determined only by the transmit beamforming vectors and the hardware-related circuit power consumption.}
The above problems (P0) and (P1) are both challenging due to the non-convex objective and constraints. 
In particular,
the presence of semi-infinite constraints in (\ref{P1_obj}), (\ref{P1_c_1}), and (\ref{P1_c_2}) 
makes (P1) more difficult to solve.
In the following, 
we will solve the above optimization problems sequentially.

\section{Solution for Perfect CSI Case}

In this section, 
we propose an efficient iterative algorithm to solve problem (P0) via invoking the FP and MM methods.

\subsection{Problem Reformulation}

First,
we note that the original rate function $\mathrm{R}_{k}$ is difficult to optimize directly 
due to the non-convex coupling between the log function and the fractional SINR term.
To facilitate the optimization of (P0),
we equivalently recast the rate function $\mathrm{R}_{k}$ 
by using the FP framework \cite{ref_FP}$-$\cite{ref_FP_1}.
Specifically,
by introducing auxiliary variables 
$\{ \gamma_{k} \}$ 
and leveraging the Lagrangian dual reformulation,
we can recast
the original rate function $\mathrm{R}_{k}$ into (\ref{FP_1}).
\begin{figure*}
\begin{align}
&\mathrm{\dot{R}}_{k}(\{\mathbf{w}_{c,k}\}, \{\mathbf{w}_{r,n}\}, \gamma_{k})
=\text{log}(1+ \gamma_{k})
-
\gamma_{k}
+
\frac{(1+\gamma_{k})\vert{\mathbf{h}}_{c,k}^H\mathbf{w}_{c,k}  \vert^2}
{
\sum_{i=1}^{K}\vert {\mathbf{h}}_{c,k}^H\mathbf{w}_{c,i}\vert^2
+
\sum_{n=1}^{N} \vert {\mathbf{h}}_{c,k}^H\mathbf{w}_{r,n}\vert^2
+
\sigma_{c,k}^2
}.\label{FP_1}
\end{align}
\boldsymbol{\hrule}
\end{figure*}
Here, each auxiliary variable 
$\gamma_{k}$ serves as a surrogate for the corresponding SINR term.
And then, 
with the aid of the auxiliary variables $\{  \omega_{k} \}$,
the function (\ref{FP_1}) can be further transformed into (\ref{FP_2}) via the quadratic transform,
\begin{figure*}
\begin{align}
\mathrm{\ddot{R}}_{k}(\{\mathbf{w}_{c,k}\}, \{\mathbf{w}_{r,n}\}, \gamma_{k}, \omega_{k})
&=\text{log}(1+ \gamma_{k})
-
\gamma_{k}
+
\bigg(
2\sqrt{(1+\gamma_{k})}\text{Re}
\{
\omega_{k}^{\ast}
{\mathbf{h}}_{c,k}^H\mathbf{w}_{c,k}
\}\label{FP_2}\\
&-
\vert \omega_{k}\vert^2
( {\sum}_{i=1}^{K}\vert {\mathbf{h}}_{c,k}^H\mathbf{w}_{c,i}\vert^2
+
{\sum}_{n=1}^{N} \vert {\mathbf{h}}_{c,k}^H\mathbf{w}_{r,n}\vert^2
+
\sigma_{c,k}^2 )
\bigg).
\nonumber
\end{align}
\boldsymbol{\hrule}
\end{figure*}
where
$\omega_{k}	$ is introduced to handle the remaining fractional term in (\ref{FP_1}) by converting it into a more tractable quadratic form.

Therefore,
using the above transformations, 
(P0) can be rewritten in an equivalent form as
\begin{subequations}
\begin{align}
&\textrm{(P2)}:\mathop{\textrm{max}}
\limits_{
\substack{
\{\mathbf{w}_{c,k}\},\{\mathbf{w}_{r,n}\},\\
\{\gamma_{k}\},
\{\omega_{k}\}
}
}\
\frac{
{\sum}_{k=1}^{K}
\mathrm{\ddot{R}}_{k}(\{\mathbf{w}_{c,k}\},\{\mathbf{w}_{r,n}\},\gamma_{k},
\omega_{k} ) }
{\mathrm{P}_{TRIS}(\{\mathbf{w}_{c,k}\},\{\mathbf{w}_{r,n}\})}
\label{P2_obj}\\
&\textrm{s.t.}\ 
\mathrm{\ddot{R}}_{k}(\{\mathbf{w}_{c,k}\},\{\mathbf{w}_{r,n}\},\gamma_{k},
\omega_{k}) \geq R_{th}, \forall k \in \mathcal{K},\label{P2_c_1}\\
&  \mathrm{P}_{target}(\{\mathbf{w}_{c,k}\},\{\mathbf{w}_{r,n}\})  \geq P_r, \label{P2_c_2}\\
& {\sum}_{k=1}^{K}
\mathbf{w}_{c,k}^H {\mathbf{A}}_{n} \mathbf{w}_{c,k} 
+ {\sum}_{i=1}^{N}\mathbf{w}_{r,i}^H {\mathbf{A}}_{n}\mathbf{w}_{r,i} 
\leq P_{unit}.  \label{P2_c_3} 
\end{align}
\end{subequations}

Note that problem (P2) is still difficult to solve due to the fractional objective (\ref{P2_obj}) 
and the non-convex constraints (\ref{P2_c_1}) and (\ref{P2_c_2}).

Therefore,
to reformulate the fractional objective into a more tractable form, 
we introduce the slack variables 
$\mathbf{x} = [x_1, \cdots ,x_K]^T \in \mathbb{R}^{K \times 1}$ and $y$
for the numerator and denominator terms, respectively,
and thereby obtain the following equivalent formulation of problem (P2),
which is given as
\begin{subequations}
\begin{align}
&\textrm{(P3)}:\mathop{\textrm{max}}
\limits_{
\substack{
\{\mathbf{w}_{c,k}\},\{\mathbf{w}_{r,n}\},\\
\{\gamma_{k}\},
\{\omega_{k}\},
\mathbf{x} \geq 0,y
}
}\
\frac{
{\sum}_{k=1}^{K}
x_k^2 }
{y}
\label{P3_obj}\\
&\textrm{s.t.}\ 
\mathrm{\ddot{R}}_{k}(\{\mathbf{w}_{c,k}\},\{\mathbf{w}_{r,n}\}, 
\gamma_{k},
\omega_{k}) \geq x_k^2, \forall k \in \mathcal{K},\label{P3_c_1}\\
& x_k^2  \geq R_{th}, \forall k \in \mathcal{K},\label{P3_c_1_1}\\
&\mathrm{P}_{TRIS}(\{\mathbf{w}_{c,k}\},\{\mathbf{w}_{r,n}\}) \leq y, \label{P3_c_1_2}\\
&\mathrm{P}_{target}(\{\mathbf{w}_{c,k}\},\{\mathbf{w}_{r,n}\})  \geq P_r, \label{P3_c_2}\\
& {\sum}_{k=1}^{K}
\mathbf{w}_{c,k}^H {\mathbf{A}}_{n} \mathbf{w}_{c,k} 
+ {\sum}_{i=1}^{N}\mathbf{w}_{r,i}^H {\mathbf{A}}_{n}\mathbf{w}_{r,i} 
\leq P_{unit}.  \label{P3_c_3} 
\end{align}
\end{subequations}

The coupling variables and the non-convex objective and constraints still render problem (P3) difficult to tackle directly.
Therefore,
we apply the block coordinate ascent (BCA) method \cite{ref_BCA} to solve problem (P3).
Concretely,
the optimization variables are split into three blocks,
i.e.,
$\{\{\mathbf{w}_{c,k}\}, \{\mathbf{w}_{r,n}\}, \mathbf{x}, y\}$,
$\{\gamma_{k}\}$
and
$ \{\omega_{k}\}$,
and the objective in (P3) is maximized by blockwise alternating optimization: 
at each iteration one block is optimized while the remaining blocks are fixed, 
and this cycle is repeated until convergence.

\subsection{Optimizing Auxiliary Variables}

With the other variables fixed, 
the auxiliary variables 
$\gamma_{k}$
and
$ \omega_{k}$ are respectively updated by
\begin{align}
&\gamma_{k}^{\star}
=\frac{\vert{\mathbf{h}}_{c,k}^H\mathbf{w}_{c,k}  \vert^2}
{
\sum_{i\neq k}^{K}\vert{\mathbf{h}}_{c,k}^H\mathbf{w}_{c,i}\vert^2
+
\sum_{n=1}^{N} \vert {\mathbf{h}}_{c,k}^H\mathbf{w}_{r,n}\vert^2
+
\sigma_{c,k}^2
},\label{FP_1_solution}\\
&\omega_{k}^{\star}
=\frac{\sqrt{1+\gamma_{k}}{\mathbf{h}}_{c,k}^H\mathbf{w}_{c,k}}
{
\sum_{i=1}^{K}\vert {\mathbf{h}}_{c,k}^H\mathbf{w}_{c,i}\vert^2
+
\sum_{n=1}^{N} \vert {\mathbf{h}}_{c,k}^H\mathbf{w}_{r,n}\vert^2
+
\sigma_{c,k}^2
}.\label{FP_2_solution}
\end{align}

\subsection{Updating the Beamformer}

In this subsection, 
we present the method to update the beamformers $\{\mathbf{w}_{c,k}\}$ and $\{ \mathbf{w}_{r,n}\}$
and the slack variables $\{\mathbf{x},y\}$.
First, 
we define the following new notations
\begin{align}
&\mathbf{b}_{1,k}
\triangleq
\sqrt{1+\gamma_{k}}(\omega_{k}{\mathbf{h}}_{c,k}),
C_{1,k}
\triangleq
\text{log}(1+\gamma_{k})-\gamma_{k}-\vert\omega_{k}\vert^2\sigma_{c,k}^2,
\mathbf{B}_{1,k}
\triangleq
\vert\omega_{k}\vert^2{\mathbf{h}}_{c,k}{\mathbf{h}}_{c,k}^H.
\end{align}

Based on the above notation, 
we can rearrange the function $\mathrm{\ddot{R}}_{k}(\mathbf{w}_c, \mathbf{w}_r, \gamma_{k}, \omega_{k})$ into 
a function with respect to (w.r.t.) the variables $\{\{\mathbf{w}_{c,k}\}, \{\mathbf{w}_{r,n}\}\}$ explicitly, 
which is formulated as
\begin{align}
&\mathrm{\ddot{R}}_{k}(\{\mathbf{w}_{c,k}\}, \{\mathbf{w}_{r,n}\}, \gamma_{k}, \omega_{k}) \label{Obj_trans}\\
&=
-{\sum}_{i=1}^{K} \mathbf{w}_{c,i}^H\mathbf{B}_{1,k}\mathbf{w}_{c,i} 
-{\sum}_{n=1}^{N} \mathbf{w}_{r,n}^H\mathbf{B}_{1,k}\mathbf{w}_{r,n}
+
2\text{Re}\{ \mathbf{b}_{1,k}^H\mathbf{w}_{c,k} \} + C_{1,k}.\nonumber
\end{align}

Therefore,
the optimization problem of $\{\{\mathbf{w}_{c,k}\}, \{\mathbf{w}_{r,n}\}, \mathbf{x},y\}$
is meant to solve the following problem
\begin{subequations}
\begin{align}
&\textrm{(P4)}:\mathop{\textrm{max}}
\limits_{\{\mathbf{w}_{c,k}\},\{\mathbf{w}_{r,n}\},\mathbf{x} \geq 0,y
}\
\frac{
{\sum}_{k=1}^{K}
x_k^2 }
{y}
\label{P4_obj}\\
&\textrm{s.t.}\ 
-{\sum}_{i=1}^{K} \mathbf{w}_{c,i}^H\mathbf{B}_{1,k}\mathbf{w}_{c,i} 
-{\sum}_{n=1}^{N} \mathbf{w}_{r,n}^H\mathbf{B}_{1,k}\mathbf{w}_{r,n} \label{P4_c_1}\\
&+
2\text{Re}\{ \mathbf{b}_{1,k}^H\mathbf{w}_{c,k} \} + C_{1,k} \geq x_k^2, \forall k \in \mathcal{K},\nonumber\\
& x_k^2  \geq R_{th}, \forall k \in \mathcal{K},\label{P4_c_1_1}\\
&\xi_{TRIS}{\sum}_{n=1}^{N}\bigg( {\sum}_{k=1}^{K}\mathbf{w}_{c,k}^H {\mathbf{A}}_{n} \mathbf{w}_{c,k}
+ {\sum}_{i=1}^{N}\mathbf{w}_{r,i}^H {\mathbf{A}}_{n} \mathbf{w}_{r,i} \bigg) +P_{c,TRIS} \leq y,\label{P4_c_1_2} \\
&{\sum}_{n=1}^{N}\mathbf{w}_{r,n}^H {\mathbf{h}}_{r} {\mathbf{h}}_{r}^H \mathbf{w}_{r,n} 
\!+\!
{\sum}_{k=1}^{K}\mathbf{w}_{c,k}^H {\mathbf{h}}_{r} {\mathbf{h}}_{r}^H \mathbf{w}_{c,k}\!  \geq\! P_r, \label{P4_c_2}\\
& {\sum}_{k=1}^{K}
\mathbf{w}_{c,k}^H {\mathbf{A}}_{n} \mathbf{w}_{c,k} 
+ {\sum}_{i=1}^{N}\mathbf{w}_{r,i}^H {\mathbf{A}}_{n}\mathbf{w}_{r,i} 
\leq P_{unit}.  \label{P4_c_3} 
\end{align}
\end{subequations}

The above problem (P4) is still difficult to solve 
due to the non-convex objective function and constraints.

First,
since the function ${x_k^2}/{y}$ is jointly convex in $(x_k, y)$ \cite{ref_EE_joint_convex},
inspired by the MM framework,
we convexify the function ${x_k^2}/{y}$ via linearization as follows
\begin{align}
\frac{x_k^2}{y}\geq\bigg( \frac{2x_{k,0}}{y_0}x_{k}-\frac{x_{k,0}^2}{y_0^2}y \bigg),
\forall k \in \mathcal{K}, \label{EE_MM_1}
\end{align}
where
$x_{k,0}$
and
$y_0$
are obtained from the last iteration.

Moreover,
the non-convex constraint (\ref{P4_c_1_1}) can be convexified by the MM method \cite{ref_MM}
as follows
\begin{align}
{x_k^2}\geq 2x_{k,0}x_{k} - x_{k,0}^2,
\forall k \in \mathcal{K}.
\end{align}

Note that
the beampattern gain constraint (\ref{P4_c_2}) is non-convex.
To tackle this difficulty,
the MM method is adopted.
Specifically,
by using the MM methodology,
a tight lower bound of the term 
$\mathbf{w}_{r,n}^H {\mathbf{h}}_{r} {\mathbf{h}}_{r}^H \mathbf{w}_{r,n} $ in (\ref{P4_c_2}) 
can be given as
\begin{align}
&\mathbf{w}_{r,n}^H {\mathbf{h}}_{r} {\mathbf{h}}_{r}^H \mathbf{w}_{r,n}
\geq
\mathbf{w}_{r,n,0}^H {\mathbf{h}}_{r} {\mathbf{h}}_{r}^H\mathbf{w}_{r,n,0}
+
2\text{Re}\{ \mathbf{w}_{r,n,0}^H {\mathbf{h}}_{r} {\mathbf{h}}_{r}^H(\mathbf{w}_{r,n} - \mathbf{w}_{r,n,0}  )  \}, \label{SOCP_MM_1}\\
&= 2\text{Re}\{ \mathbf{w}_{r,n,0}^H {\mathbf{h}}_{r} {\mathbf{h}}_{r}^H\mathbf{w}_{r,n}\}+C_{2,n},\nonumber
\end{align}
where
$\mathbf{w}_{r,n,0}$ is the value obtained from the last iteration
and $C_{2,n}$ is constant.

Similarly,
we can obtain a tight lower bound of the term 
$\mathbf{w}_{c,k}^H {\mathbf{h}}_{r} {\mathbf{h}}_{r}^H \mathbf{w}_{c,k} $
in (\ref{P4_c_2}) 
by invoking the MM method again,
which can be given as follows
\begin{align}
&\mathbf{w}_{c,k}^H {\mathbf{h}}_{r} {\mathbf{h}}_{r}^H \mathbf{w}_{c,k}
\geq
\mathbf{w}_{c,k,0}^H {\mathbf{h}}_{r} {\mathbf{h}}_{r}^H\mathbf{w}_{c,k,0}
+
2\text{Re}\{ \mathbf{w}_{c,k,0}^H {\mathbf{h}}_{r} {\mathbf{h}}_{r}^H(\mathbf{w}_{c,k} - \mathbf{w}_{c,k,0}  )  \}, \label{SOCP_MM_1}\\
&= 2\text{Re}\{ \mathbf{w}_{c,k,0}^H {\mathbf{h}}_{r} {\mathbf{h}}_{r}^H\mathbf{w}_{c,k}\}+C_{3,k},\nonumber
\end{align}
with $\mathbf{w}_{c,k,0}$ being obtained from the last iteration
and $C_{3,k}$ being a constant term.

Therefore,
the problem (P4) can be rewritten as
\begin{subequations}
\begin{align}
&\textrm{(P5)}:\mathop{\textrm{max}}
\limits_{\{\mathbf{w}_{c,k}\},\{\mathbf{w}_{r,n}\},\mathbf{x}\geq 0,y
}\
{\sum}_{k=1}^{K}
\bigg( \frac{2x_{k,0}}{y_0}x_{k}\!-\!\frac{x_{k,0}^2}{y_0^2}y \bigg)
\label{P5_obj}\\
&\textrm{s.t.}\ 
-{\sum}_{i=1}^{K} \mathbf{w}_{c,i}^H\mathbf{B}_{1,k}\mathbf{w}_{c,i} 
-{\sum}_{n=1}^{N} \mathbf{w}_{r,n}^H\mathbf{B}_{1,k}\mathbf{w}_{r,n} \label{P5_c_1}\\
&+
2\text{Re}\{ \mathbf{b}_{1,k}^H\mathbf{w}_{c,k} \} + C_{1,k} \geq x_k^2, \forall k \in \mathcal{K},\nonumber\\
& 2x_{k,0}x_{k} - x_{k,0}^2  \geq R_{th}, \forall k \in \mathcal{K},\label{P5_c_1_1}\\
&\xi_{TRIS}{\sum}_{n=1}^{N}\bigg( {\sum}_{k=1}^{K}\mathbf{w}_{c,k}^H {\mathbf{A}}_{n} \mathbf{w}_{c,k} 
+ {\sum}_{i=1}^{N}\mathbf{w}_{r,i}^H {\mathbf{A}}_{n} \mathbf{w}_{r,i} \bigg) +P_{c,TRIS} \leq y,\label{P5_c_1_2}\\
&{\sum}_{n=1}^{N}(
2\text{Re}\{ \mathbf{w}_{r,n,0}^H {\mathbf{h}}_{r} {\mathbf{h}}_{r}^H\mathbf{w}_{r,n} \} + C_{2,n}  )\label{P5_c_2}\\
&+
{\sum}_{k=1}^{K}
(
2\text{Re}\{ \mathbf{w}_{c,k,0}^H {\mathbf{h}}_{r} {\mathbf{h}}_{r}^H\mathbf{w}_{c,k}\}+ C_{3,k})  \geq P_r, \nonumber\\
& {\sum}_{k=1}^{K}
\mathbf{w}_{c,k}^H {\mathbf{A}}_{n} \mathbf{w}_{c,k} 
+ {\sum}_{i=1}^{N}\mathbf{w}_{r,i}^H {\mathbf{A}}_{n}\mathbf{w}_{r,i} 
\leq P_{unit}.  \label{P5_c_3} 
\end{align}
\end{subequations}

The problem (P5) is convex and can be solved by CVX \cite{ref_CVX}.
The algorithm to solve the EE maximization problem, i.e., (P0), is specified in Algorithm \ref{alg:1}.

\begin{algorithm}[t]
\caption{Solving the Problem (P0)}
\label{alg:1}
\begin{algorithmic}[1]
\STATE {initialize}
$\{\mathbf{w}_{c,k}^{(0)}\}$,
$\{\mathbf{w}_{r,n}^{(0)}\}$,
$\mathbf{x}^{(0)}$,
$y^{(0)}$,
and
$t=0$
;
\REPEAT
\STATE update $\{\gamma_k^{(t+1)}\}$ and $\{\omega_k^{(t+1)}\}$ by (\ref{FP_1_solution}) and (\ref{FP_2_solution}), respectively;
\STATE update $\{\mathbf{w}_{c,k}^{(t+1)}\}$,  $\{\mathbf{w}_{r,n}^{(t+1)}\}$,
 $\mathbf{x}^{(t+1)}$, and $y^{(t+1)}$ by solving  (P5);
\STATE $t++$;
\UNTIL{$convergence$;}
\end{algorithmic}
\end{algorithm}

\subsection{Convergence and Complexity Analyses}
\textit{ 1) Convergence Analysis:}
Algorithm \ref{alg:1} 
guarantees a feasible solution and monotonic convergence of the objective function, 
as established by the following theorem, 
whose proof can be found in Appendix A:
\begin{theorem} \label{theorem_1}
When Algorithm \ref{alg:1} starts from a feasible point,
in each iteration,
the solution produced by Algorithm \ref{alg:1} is feasible for problem (P5),
and the objective value for problem (P5) increases monotonically. 
\end{theorem}

\textit{ 2) Complexity Analysis:} 
{
In Algorithm~\ref{alg:1}, 
the auxiliary variables $\{\gamma_k\}$ and $\{\omega_k\}$ are updated in closed form according to (\ref{FP_1_solution}) and (\ref{FP_2_solution}), respectively. 
Therefore, 
the computational complexity of Algorithm~\ref{alg:1} is mainly dominated by solving the problem (P5).
The complexity of solving (P5) is 
$
C_{\mathrm{P5}}
=
\mathcal{O}\left(
\sqrt{3K+N+2}
\left(2N(K+N)+K+1\right)^3
\log\left(\frac{1}{\epsilon}\right)
\right)$.
Let $C_1$ denote the number of outer iterations required by Algorithm~\ref{alg:1}. 
Then, the overall complexity of Algorithm~\ref{alg:1}
is
${C}_{\mathrm{Alg.1}}
=C_1C_{\mathrm{P5}}
$.
By omitting the logarithmic accuracy factor and lower-order terms, 
the complexity can be further approximated as
${C}_{\mathrm{Alg.1}}
=\mathcal{O}\left(
C_1N^{3.5}(K+N)^3
\right)$.
}

\section{Solution for Imperfect CSI Case}

In this section, 
an algorithm is developed based on the S-Procedure and MM method for solving the worst-case problem (P1).

\subsection{Updating The Beamformer}

First,
by introducing the slack variables 
$\bar{\mathbf{x}} = [\bar{x}_1, \cdots ,\bar{x}_K]^T \in \mathbb{R}^{K \times 1}$ and $\bar{y}$,
the problem (P1) can be rewritten as
\begin{subequations}
\begin{align}
&\textrm{(P6)}:\mathop{\textrm{max}}
\limits_{\{\mathbf{w}_{c,k}\},\{\mathbf{w}_{r,n}\},\bar{\mathbf{x}} \geq \mathbf{0}, \bar{y}
}\
\frac{ {\sum}_{k=1}^{K} \bar{x}_k^2 } {\bar{y}}
\label{P6_obj}\\
&\textrm{s.t.}\ \mathrm{R}_{k}(\{\mathbf{w}_{c,k}\},\{\mathbf{w}_{r,n}\}) \geq \bar{x}_k^2, 
\forall \mathbf{h}_{c,k} \in \mathcal{H}_{c,k}, 
\forall k \in \mathcal{K},\label{P6_c_1}\\
& \bar{x}_k^2 \geq R_{th}, \forall k \in \mathcal{K},\label{P6_c_2}\\
&\xi_{TRIS}{\sum}_{n=1}^{N}\bigg( {\sum}_{k=1}^{K}\mathbf{w}_{c,k}^H {\mathbf{A}}_{n} \mathbf{w}_{c,k}
+ {\sum}_{i=1}^{N}\mathbf{w}_{r,i}^H {\mathbf{A}}_{n} \mathbf{w}_{r,i} \bigg) +P_{c,TRIS} \leq \bar{y}, \label{P6_c_3}\\
&  {\sum}_{i=1}^{N}\vert {\mathbf{h}}_{r}^H \mathbf{w}_{r,i} \vert^2
+{\sum}_{k=1}^{K}\vert {\mathbf{h}}_{r}^H \mathbf{w}_{c,k}\vert^2  \geq P_r, 
\forall \mathbf{h}_{r} \in\mathcal{H}_{r}, \label{P6_c_4}\\
& {\sum}_{k=1}^{K}
\mathbf{w}_{c,k}^H {\mathbf{A}}_{n} \mathbf{w}_{c,k} 
+ {\sum}_{i=1}^{N}\mathbf{w}_{r,i}^H {\mathbf{A}}_{n}\mathbf{w}_{r,i} 
\leq P_{unit}.  \label{P6_c_5}
\end{align}
\end{subequations}

And then,
we reformulate the rate function 
$\mathrm{R}_{k}(\{\mathbf{w}_{c,k}\},\{\mathbf{w}_{r,n}\})$ 
as follows
\begin{align}
&\mathrm{R}_{k}(\{\mathbf{w}_{c,k}\},\{\mathbf{w}_{r,n}\})
=\text{log}\bigg(1+\frac{\vert{\mathbf{h}}_{c,k}^H\mathbf{w}_{c,k}  \vert^2}
{\sum_{i\neq k}^{K}\vert {\mathbf{h}}_{c,k}^H\mathbf{w}_{c,i}\vert^2
+\sum_{n=1}^{N} \vert {\mathbf{h}}_{c,k}^H\mathbf{w}_{r,n}\vert^2
+
\sigma_{c,k}^2}\bigg)\label{R_equ_trans_1}\\
&=\underbrace{
\text{log}\bigg({\sum}_{i=1}^{K}\vert {\mathbf{h}}_{c,k}^H\mathbf{w}_{c,i}\vert^2
+{\sum}_{n=1}^{N} \vert {\mathbf{h}}_{c,k}^H\mathbf{w}_{r,n}\vert^2
+
\sigma_{c,k}^2\bigg)}
\limits_{\mathrm{\hat{R}}_{k}(\{\mathbf{w}_{c,k}\},\{\mathbf{w}_{r,n}\})}
\nonumber\\
&-
\underbrace{
\text{log}\bigg({\sum}_{i\neq k}^{K}\vert {\mathbf{h}}_{c,k}^H\mathbf{w}_{c,i}\vert^2
+{\sum}_{n=1}^{N} \vert {\mathbf{h}}_{c,k}^H\mathbf{w}_{r,n}\vert^2
+
\sigma_{c,k}^2\bigg)}
\limits_{\mathrm{\check{R}}_{k}(\{\mathbf{w}_{c,k}\},\{\mathbf{w}_{r,n}\})}
.\nonumber
\end{align}

Next,
to facilitate the subsequent robust optimization,
we introduce the slack variables
$\boldsymbol{\mu}=[\mu_1,\cdots,\mu_K]^T \in \mathbb{R}^{K \times 1}$
and
$\boldsymbol{\nu}=[\nu_1,\cdots,\nu_K]^T \in \mathbb{R}^{K \times 1}$
to decouple the two composite terms in (\ref{R_equ_trans_1}).
Accordingly,
(\ref{R_equ_trans_1}) can be equivalently reformulated as follows
\begin{align}
&\mathrm{\hat{R}}_{k}(\{\mathbf{w}_{c,k}\},\{\mathbf{w}_{r,n}\})
-
\mathrm{\check{R}}_{k}(\{\mathbf{w}_{c,k}\},\{\mathbf{w}_{r,n}\})
\geq \bar{x}_k^2\\
&\Leftrightarrow
\begin{cases}
\text{log}(\mu_k)
-\text{log}(\nu_k)
\geq \bar{x}_k^2, \\
{\sum}_{i=1}^{K}\vert {\mathbf{h}}_{c,k}^H\mathbf{w}_{c,i}\vert^2
+{\sum}_{n=1}^{N} \vert {\mathbf{h}}_{c,k}^H\mathbf{w}_{r,n}\vert^2
+
\sigma_{c,k}^2 \geq \mu_k,\\
{\sum}_{i\neq k}^{K}\vert {\mathbf{h}}_{c,k}^H\mathbf{w}_{c,i}\vert^2
+{\sum}_{n=1}^{N} \vert {\mathbf{h}}_{c,k}^H\mathbf{w}_{r,n}\vert^2
+
\sigma_{c,k}^2\leq \nu_k.
\end{cases}\nonumber
\end{align}

Therefore,
(P6) can be transformed into the following form
\begin{subequations}
\begin{align}
&\textrm{(P7)}:\mathop{\textrm{max}}
\limits_{
\substack{\{\mathbf{w}_{c,k}\},\{\mathbf{w}_{r,n}\},\\
\bar{\mathbf{x}},
\bar{y},
\boldsymbol{\mu},
\boldsymbol{\nu}
}
}\
\frac{ {\sum}_{k=1}^{K} \bar{x}_k^2 } {\bar{y}}
\label{P7_obj}\\
&\textrm{s.t.}\ 
\text{log}(\mu_k)-\text{log}(\nu_k)\geq \bar{x}_k^2, \forall k \in \mathcal{K},\label{P7_c_1}\\
& {\sum}_{i=1}^{K}\vert {\mathbf{h}}_{c,k}^H\mathbf{w}_{c,i}\vert^2
+{\sum}_{n=1}^{N} \vert {\mathbf{h}}_{c,k}^H\mathbf{w}_{r,n}\vert^2
+
\sigma_{c,k}^2 \geq \mu_k, 
\forall \mathbf{h}_{c,k} \in \mathcal{H}_{c,k}, \forall k \in \mathcal{K},\label{P7_c_2}\\
&{\sum}_{i\neq k}^{K}\vert {\mathbf{h}}_{c,k}^H\mathbf{w}_{c,i}\vert^2
+{\sum}_{n=1}^{N} \vert {\mathbf{h}}_{c,k}^H\mathbf{w}_{r,n}\vert^2
+
\sigma_{c,k}^2\leq \nu_k, 
\forall \mathbf{h}_{c,k} \in \mathcal{H}_{c,k},\forall k \in \mathcal{K},\label{P7_c_3}\\
& \bar{x}_k^2 \geq R_{th}, \forall k \in \mathcal{K},\label{P7_c_4}\\
&\xi_{TRIS}{\sum}_{n=1}^{N}\bigg( {\sum}_{k=1}^{K}\mathbf{w}_{c,k}^H {\mathbf{A}}_{n} \mathbf{w}_{c,k} 
+ {\sum}_{i=1}^{N}\mathbf{w}_{r,i}^H {\mathbf{A}}_{n} \mathbf{w}_{r,i} \bigg) +P_{c,TRIS} \leq \bar{y},\label{P7_c_5}\\
&  {\sum}_{i=1}^{N}\vert {\mathbf{h}}_{r}^H \mathbf{w}_{r,i} \vert^2
+{\sum}_{k=1}^{K}\vert {\mathbf{h}}_{r}^H \mathbf{w}_{c,k}\vert^2  \geq P_r, 
\forall \mathbf{h}_{r} \in\mathcal{H}_{r}, \label{P7_c_6}\\
& {\sum}_{k=1}^{K}
\mathbf{w}_{c,k}^H {\mathbf{A}}_{n} \mathbf{w}_{c,k} 
+ {\sum}_{i=1}^{N}\mathbf{w}_{r,i}^H {\mathbf{A}}_{n}\mathbf{w}_{r,i} 
\leq P_{unit}, \label{P7_c_7}\\
&
\bar{\mathbf{x}}\geq 0,
\boldsymbol{\mu} \geq 0,
\boldsymbol{\nu} \geq 0.
\end{align}
\end{subequations}

Next,
we sequentially transform the non-convex semi-infinite inequality constraints, 
i.e., (\ref{P7_c_2}), (\ref{P7_c_3}), and (\ref{P7_c_6}).

First,
we introduce the following lemma.
\begin{lemma}
(General S-Procedure \cite{ref_S_Procedure}):
Considering the following quadratic functions w.r.t. the variable $\mathbf{x}\in\mathbb{C}^{N\times 1}$:
\begin{align}
f_i(\mathbf{x})=\mathbf{x}^H\mathbf{Q}_i\mathbf{x}+2\text{Re}\{\mathbf{q}_i^H\mathbf{x}\}+q_i,\quad i=0,\ldots,P,
\end{align}
with Hermitian matrices $\mathbf{Q}_i=\mathbf{Q}_i^H  \in\mathbb{C}^{N\times N}$.
The semi-infinite condition that $f_0(\mathbf{x})\ge 0$ for the variable $\mathbf{x}$ 
satisfying $f_i(\mathbf{x})\ge 0$ $(i=1,\ldots,P)$ 
is equivalent to the existence of multipliers ${\varpi_i},i=1, \cdots,P$ with $\varpi_i\ge 0$ such that
\begin{equation}
\begin{bmatrix}
\mathbf{Q}_0 & \mathbf{q}_0\\
\mathbf{q}_0^H & q_0
\end{bmatrix}
-\sum_{i=1}^{P}\varpi_i
\begin{bmatrix}
\mathbf{Q}_i & \mathbf{q}_i\\
\mathbf{q}_i^H & q_i
\end{bmatrix}
\succeq \mathbf{0}.
\end{equation}
\end{lemma}

Furthermore,
by leveraging the MM method and the S-Procedure,
the constraint (\ref{P7_c_2}) can be transformed as
\begin{align}
&{\sum}_{i=1}^{K}\vert {\mathbf{h}}_{c,k}^H\mathbf{w}_{c,i}\vert^2
+{\sum}_{n=1}^{N} \vert {\mathbf{h}}_{c,k}^H\mathbf{w}_{r,n}\vert^2
+
\sigma_{c,k}^2 \geq \mu_k\\
&\Rightarrow
\left[
\begin{array}{cc}
\varpi_{1,k} \mathbf{I} + \mathbf{B}_{9} & \mathbf{b}_{11,k} \\
\mathbf{b}_{11,k}^H  & c_{9,k}+\sigma_{c,k}^2 - \mu_k - \varpi_{1,k}\xi_{c,k}^2
\end{array}
\right] \succeq \mathbf{0},\nonumber
\end{align}
where the detailed derivation and the newly introduced coefficients are provided in Appendix B.

Next,
we introduce the following lemma.
\begin{lemma}
(General sign-definiteness \cite{ref_sign}):
Let $\mathbf{Q}=\mathbf{Q}^H$ and $\{\mathbf{D}_i,\mathbf{M}_i\}_{i=1}^{P}$ be given.
Then the semi-infinite LMI
\begin{equation}
\mathbf{Q}\succeq\! {\sum}_{i=1}^{P}(\mathbf{D}_i^{H}\mathbf{X}_i\mathbf{M}_i+\mathbf{M}_i^{H}\mathbf{X}_i^{H}\mathbf{D}_i), \forall i, \|\mathbf{X}_i\|_{F}\leq \xi_i,
\end{equation}
 is satisfied if and only if there exist nonnegative real numbers
 $\mu_i$, $i=1,\cdots,P$, such that
 \begin{equation}
\begin{bmatrix}
\mathbf{Q}-\sum_{i=1}^{P}\mu_i\,\mathbf{M}_i^{H}\mathbf{M}_i
& -\xi_{1}\mathbf{D}_{1}^{H} & \cdots & -\xi_{P}\mathbf{D}_{P}^{H} \\
-\xi_{1}\mathbf{D}_{1} & \mu_{1}\mathbf{I} & \cdots & \mathbf{0} \\
\vdots & \vdots & \ddots & \vdots \\
-\xi_{P}\mathbf{D}_{P} & \mathbf{0} & \cdots & \mu_{P}\mathbf{I}
\end{bmatrix}
\succeq \mathbf{0}.
\end{equation}
\end{lemma}

By adopting Schur's complement Lemma \cite{ref_Convex Optimization}
and the sign-definiteness lemma,
the constraint (\ref{P7_c_3}) can be recast as follows
\begin{align}
&{\sum}_{i\neq k}^{K}\vert {\mathbf{h}}_{c,k}^H\mathbf{w}_{c,i}\vert^2
+{\sum}_{n=1}^{N} \vert {\mathbf{h}}_{c,k}^H\mathbf{w}_{r,n}\vert^2
+\sigma_{c,k}^2\leq \nu_k \\
&\Rightarrow
\left[
\begin{array}{ccc}
\nu_k - \sigma_{c,k}^2-\varpi_{2,k} & \hat{\mathbf{h}}_{c,k}^H \mathbf{W}_{cr,-k} & \mathbf{0}\\
\mathbf{W}_{cr,-k}^H\hat{\mathbf{h}}_{c,k}  &  \mathbf{I}  & \xi_{c,k}\mathbf{W}_{cr,-k}^H\\
\mathbf{0}  &  \xi_{c,k}\mathbf{W}_{cr,-k}  & \varpi_{2,k}\mathbf{I}
\end{array}
\right] 
\succeq 
\mathbf{0},
\nonumber
\end{align}
where the detailed derivation and the newly introduced notation are given in Appendix C.

Similar to (\ref{P7_c_2}), 
by using the MM method and the S-Procedure,
the constraint (\ref{P7_c_6}) can be reformulated as follows
\begin{align}
&{\sum}_{i=1}^{N}\vert {\mathbf{h}}_{r}^H \mathbf{w}_{r,i} \vert^2
+{\sum}_{k=1}^{K}\vert {\mathbf{h}}_{r}^H \mathbf{w}_{c,k}\vert^2  \geq P_r\\
&\Rightarrow
\left[
\begin{array}{cc}
\varpi_{3} \mathbf{I} + \mathbf{B}_{18} & \mathbf{b}_{22} \\
\mathbf{b}_{22}^H  & c_{18}-P_r - \varpi_{3}\xi_{r}^2
\end{array}
\right] \succeq \mathbf{0},\nonumber
\end{align}
where the full derivation and the new parameters are provided in Appendix D.

We note that the objective (\ref{P7_obj}) and constraints (\ref{P7_c_1}) and (\ref{P7_c_4}) are still non-convex.
By leveraging the MM method,
(\ref{P7_obj}), (\ref{P7_c_1}), and (\ref{P7_c_4})
can be approximated as follows
\begin{align}
&\frac{\bar{x}_k^2}{y}\geq\bigg( \frac{2\bar{x}_{k,0}}{\bar{y}_0}\bar{x}_{k}-\frac{\bar{x}_{k,0}^2}{\bar{y}_0^2}\bar{y} \bigg),
\forall k \in \mathcal{K},\\
&\text{log}(\nu_k) \leq 
\text{log}(\nu_{k,0}) + \frac{1}{\nu_{k,0}}(\nu_{k}-\nu_{k,0}), \forall k \in \mathcal{K},\\
&{\bar{x}_k^2}\geq 2\bar{x}_{k,0}\bar{x}_{k} - \bar{x}_{k,0}^2,
\forall k \in \mathcal{K},
\end{align}
respectively,
where
$\bar{x}_k$,
$\bar{y}_0$,
and
$\nu_{k,0}$ are obtained from the last iteration.

Therefore,
based on the above transformations,
the update of the beamformers $\{\mathbf{w}_{c,k}\}$ and $\{ \mathbf{w}_{r,n}\}$
is intended to solve the following problem
\begin{subequations}
\begin{align}
&\textrm{(P8)}:\mathop{\textrm{max}}
\limits_{\substack{\{\mathbf{w}_{c,k}\},\{\mathbf{w}_{r,n}\},\bar{\mathbf{x}},\\ 
\bar{y},
\boldsymbol{\mu},
\boldsymbol{\nu},
\boldsymbol{\varpi}_1,
\boldsymbol{\varpi}_2,
{\varpi}_3
}}\
{\sum}_{k=1}^{K}
\bigg( \frac{2\bar{x}_{k,0}}{\bar{y}_0}\bar{x}_{k}-\frac{\bar{x}_{k,0}^2}{\bar{y}_0^2}\bar{y} \bigg)
\label{P8_obj}\\
&\textrm{s.t.}\ 
\text{log}(\mu_k)-
\bigg(\text{log}(\nu_{k,0}) + \frac{1}{\nu_{k,0}}(\nu_{k}-\nu_{k,0})\bigg)
\geq \bar{x}_k^2,\label{P8_c_1}\\
&\left[
\begin{array}{cc}
\varpi_{1,k} \mathbf{I} + \mathbf{B}_{9} & \mathbf{b}_{11,k} \\
\mathbf{b}_{11,k}^H  & c_{9,k}+\sigma_{c,k}^2 - \mu_k - \varpi_{1,k}\xi_{c,k}^2
\end{array}
\right] \succeq \mathbf{0},\label{P8_c_2}\\
&\left[
\begin{array}{ccc}
\nu_k - \sigma_{c,k}^2-\varpi_{2,k} & \hat{\mathbf{h}}_{c,k}^H \mathbf{W}_{cr,-k} & \mathbf{0}\\
\mathbf{W}_{cr,-k}^H\hat{\mathbf{h}}_{c,k}  &  \mathbf{I}  & \xi_{c,k}\mathbf{W}_{cr,-k}^H\\
\mathbf{0}  &  \xi_{c,k}\mathbf{W}_{cr,-k}  & \varpi_{2,k}\mathbf{I}
\end{array}
\right] 
\succeq 
\mathbf{0},
\label{P8_c_3}\\
& 2\bar{x}_{k,0}\bar{x}_{k} - \bar{x}_{k,0}^2 \geq R_{th}, \forall k \in \mathcal{K},\label{P8_c_4}\\
&\xi_{TRIS}{\sum}_{n=1}^{N}\bigg( {\sum}_{k=1}^{K}\mathbf{w}_{c,k}^H {\mathbf{A}}_{n} \mathbf{w}_{c,k} 
+ {\sum}_{i=1}^{N}\mathbf{w}_{r,i}^H {\mathbf{A}}_{n} \mathbf{w}_{r,i} \bigg) +P_{c,TRIS} \leq \bar{y},\label{P8_c_5}\\
&  \left[
\begin{array}{cc}
\varpi_{3} \mathbf{I} + \mathbf{B}_{18} & \mathbf{b}_{22} \\
\mathbf{b}_{22}^H  & c_{18}-P_r - \varpi_{3}\xi_{r}^2
\end{array}
\right] \succeq \mathbf{0}, \label{P8_c_6}\\
& {\sum}_{k=1}^{K}
\mathbf{w}_{c,k}^H {\mathbf{A}}_{n} \mathbf{w}_{c,k} 
+ {\sum}_{i=1}^{N}\mathbf{w}_{r,i}^H {\mathbf{A}}_{n}\mathbf{w}_{r,i} 
\leq P_{unit}, \label{P8_c_7}\\
&
\bar{\mathbf{x}}\geq 0,
\boldsymbol{\mu} \geq 0,
\boldsymbol{\nu} \geq 0,
\boldsymbol{\varpi}_1 \geq 0,
\boldsymbol{\varpi}_2 \geq 0,
{\varpi}_3 \geq 0.
\end{align}
\end{subequations}

Obviously, 
both the objective and the constraints of problem (P8) have been convexified. 
Therefore, (P8) is convex and can be solved by using CVX.

\begin{algorithm}[t]
\caption{Solving the Problem (P1)}
\label{alg:2}
\begin{algorithmic}[1]
\STATE {initialize}
$\{\mathbf{w}_{c,k}^{(0)}\}$,
$\{\mathbf{w}_{r,n}^{(0)}\}$,
$\bar{\mathbf{x}}^{(0)}$,
$\bar{y}^{(0)}$,
$\boldsymbol{\nu}^{(0)}$,
and
$t=0$
;
\REPEAT
\STATE update
$\{\mathbf{w}_{c,k}^{(t+1)}\}$,
$\{\mathbf{w}_{r,n}^{(t+1)}\}$,
$\bar{\mathbf{x}}^{(t+1)}$,
$\bar{y}^{(t+1)}$,
$\boldsymbol{\mu}^{(t+1)}$,
$\boldsymbol{\nu}^{(t+1)}$,
$\boldsymbol{\varpi}_{1}^{(t+1)}$,
$\boldsymbol{\varpi}_{2}^{(t+1)}$,
and
$\varpi_{3}^{(t+1)}$
by solving (P8);
\STATE $t++$;
\UNTIL{$convergence$;}
\end{algorithmic}
\end{algorithm}

\subsection{Convergence and Complexity Analyses}
\textit{ 1) Convergence Analysis:}
Since the solution to (P1) employs optimization techniques similar to those used for (P0), 
the convergence analysis is omitted here for brevity and follows that for (P0).

\textit{ 2) Complexity Analysis:}
{
In Algorithm~\ref{alg:2}, 
the computational complexity is mainly dominated by solving problem (P8) in each outer iteration.
According to the complexity analysis in \cite{ref_complexity}, 
the complexity of solving (P8) is
$C_{\mathrm{P8}}
=
\mathcal{O}\left(
\sqrt{D_1}
\left(
V_2^3+V_2^2D_2+V_2D_3
\right)
\log\left(\frac{1}{\epsilon}\right)
\right)$,
where 
$V_2=2N(K+N)+5K+2$,
$D_1 = K(N+1)+K(K+2N)+(N+1)$, 
$D_2 = K(N+1)^2+K(K+2N)^2+(N+1)^2$,
and
$D_3 = K(N+1)^3+K(K+2N)^3+(N+1)^3$.
Let $C_2$ denote the number of outer iterations required by Algorithm~\ref{alg:2}. 
Then, the overall complexity of Algorithm~\ref{alg:2} is
$
C_{\mathrm{Alg.2}}
=C_2C_{\mathrm{P8}}$.
Furthermore, 
the complexity can be approximated as
$
C_{\mathrm{Alg.2}}
=
\mathcal{O}\left(
C_2K\sqrt{K(K+N)+N}N^2(K+N)^4
\right)$.}

\section{Numerical Results}

\begin{figure}[t]
	\centering
	\includegraphics[width=.40\textwidth]{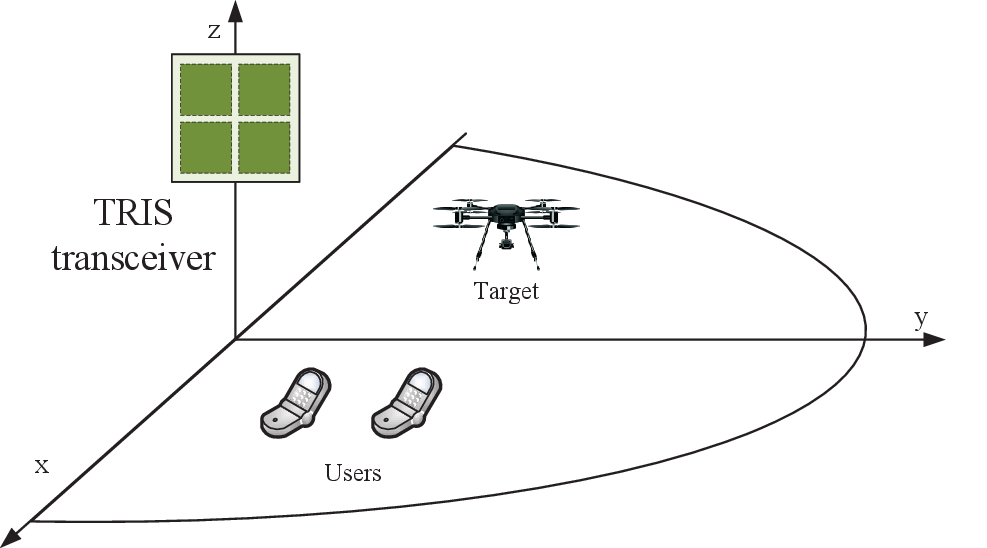}
	\caption{The considered simulation scenario.}
	\label{fig.2}
\end{figure}

In this section, 
we evaluate the performance of the proposed TRIS transceiver-enabled ISAC system 
and the effectiveness of the developed algorithms.
As shown in Fig.~\ref{fig.2}, 
a TRIS transceiver is located at $(0,0,4.5)$~m in the three-dimensional (3D) space.
The ISAC setup consists of $K=2$ users and one point-like sensing target, 
whose locations are randomly generated within a sector region with distances ranging from $20$ to $50$~m from the origin, and all nodes are placed at a height of $1.5$~m.
The path-loss exponents for the TRIS transceiver-user and TRIS transceiver-target links are set to $3.2$ and $2.2$, respectively.
The maximum transmit power at the TRIS transceiver is limited to $10$~dBm.
The noise power at each user is set to $-90$~dBm.
{
All numerical simulations are implemented in MATLAB R2023b on a PC equipped with a 2.10 GHz i7-14700 CPU and 32 GB RAM. 
}

\subsection{Perfect CSI Case}

In this subsection, 
we present simulation results on the EE of the ISAC system under perfect CSI.

\begin{figure}[t]
    \centering
    \begin{minipage}[t]{0.45\textwidth}
        \centering
        \includegraphics[width=\linewidth]{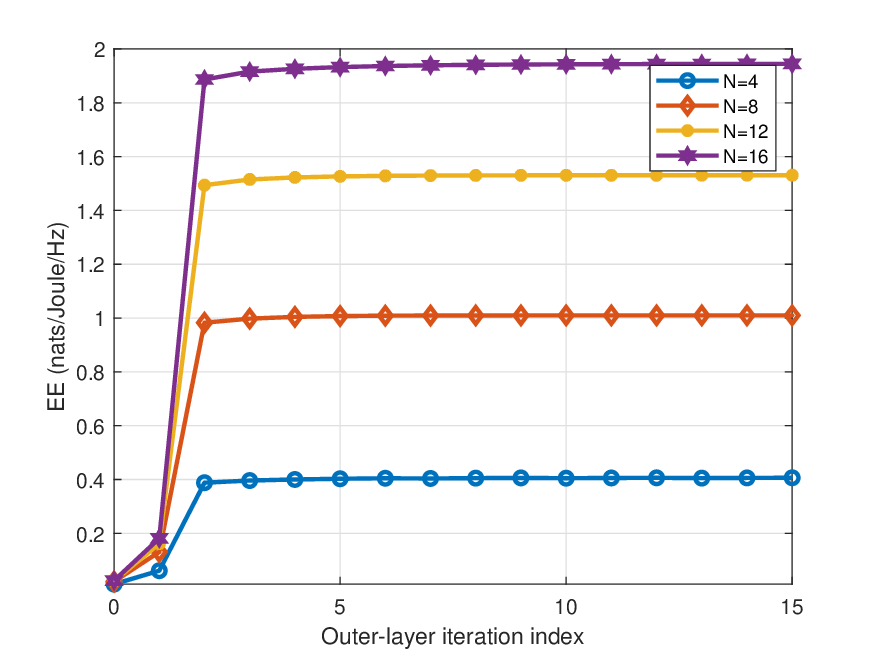}
        \captionof{figure}{{Convergence of Algorithm 1.}}
        \label{fig.3}
    \end{minipage}
    \hfill
    \begin{minipage}[t]{0.45\textwidth}
        \centering
        \includegraphics[width=\linewidth]{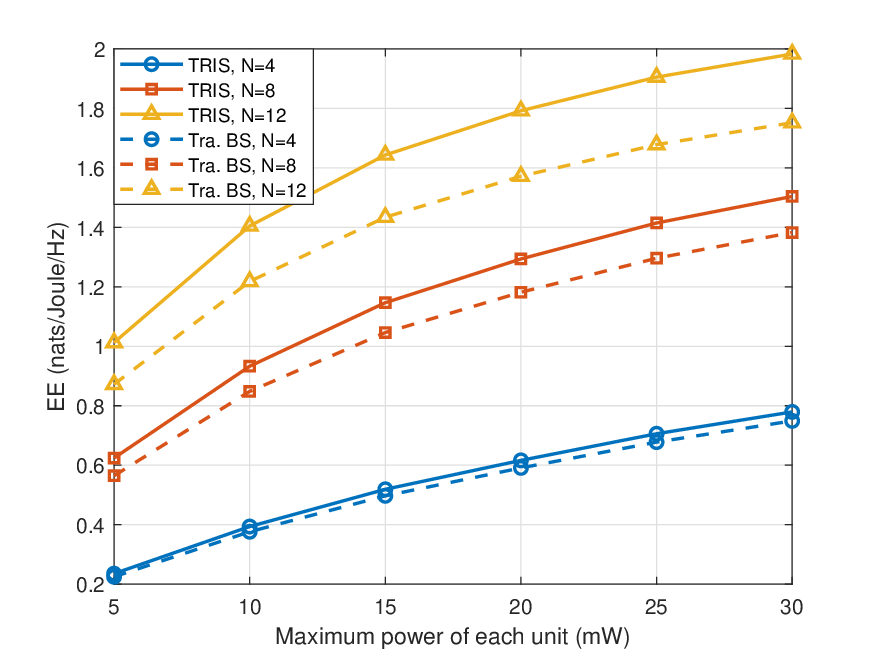}
        \captionof{figure}{EE versus the maximum power of each TRIS unit.}
        \label{fig.4}
    \end{minipage}
\end{figure}

\begin{table}[t]
\centering
\caption{{Execution Time of Algorithm~\ref{alg:1} (s)}}
\label{tab:runtime_alg_1}
\renewcommand{\arraystretch}{1.00}
\begin{tabular}{|c|c|c|c|c|}
\hline
\textbf{Algorithm}           & $N=4$  & $N=8$   & $N=12$  & $N=16$   \\
\hline
Algorithm~\ref{alg:1}     & 8.9525 & 13.2071 & 18.2107 & 31.3024 \\
\hline
\end{tabular}
\end{table}

{
Fig. \ref{fig.3} illustrates the convergence behavior of the developed algorithm for solving the EE maximization problem (P0) under different numbers of TRIS units.
Table~\ref{tab:runtime_alg_1} reports the average execution time of Algorithm~\ref{alg:1} under different values of $N$.
From Fig. \ref{fig.3},
we can observe that the objective value monotonically increases as the algorithm iterates and achieves significant improvement within the first five iterations.
Moreover, it is also shown that a larger number of TRIS elements further improves EE.}

{
For comparison, 
we consider a benchmark scheme, 
namely a conventional multi-antenna transceiver, 
referred to as ``Tra. BS'', 
which is equipped with $N$ RF chains.
Fig. \ref{fig.4} illustrates the EE versus the maximum transmit power of each transmit unit under different system configurations. 
It can be observed that the EE of all schemes increases monotonically with the maximum transmit power. 
As the maximum transmit power further increases, 
the EE curves gradually become saturated. 
Moreover, 
increasing the number of transmit units improves the EE due to the enhanced spatial beamforming degrees of freedom. 
Compared with the conventional fully-active multi-antenna transceiver, 
the proposed TRIS transceiver achieves a higher EE under all considered configurations. 
This is because the TRIS transceiver requires only a single RF chain, 
whereas the multi-antenna transceiver employs $N$ RF chains, 
resulting in higher circuit power consumption.}

\begin{figure}[t]
    \centering
    \begin{minipage}[t]{0.45\textwidth}
        \centering
        \includegraphics[width=\linewidth]{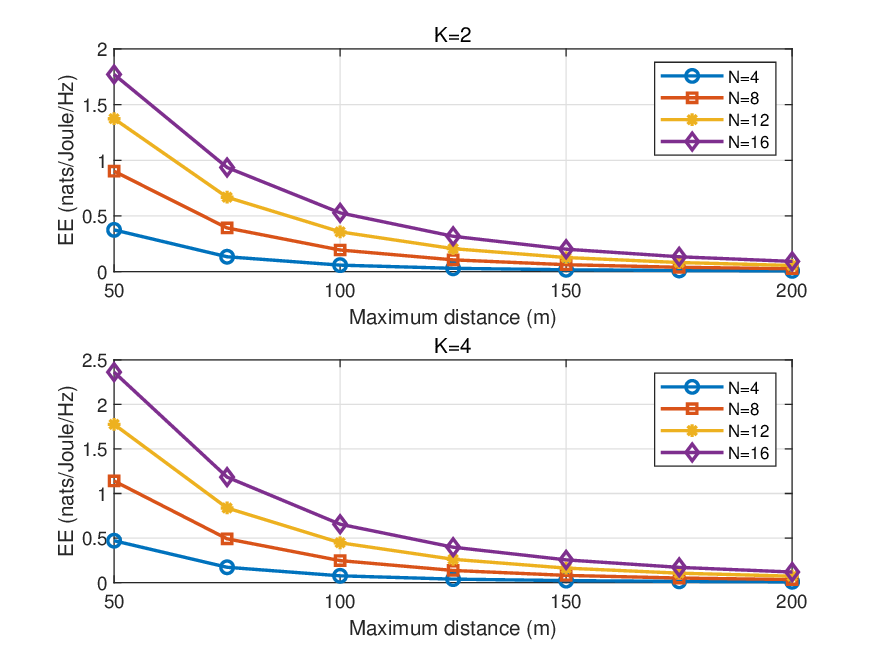}
        \captionof{figure}{EE versus the maximum distance between TRIS transceiver and users.}
        \label{fig.5}
    \end{minipage}
    \hfill
    \begin{minipage}[t]{0.45\textwidth}
        \centering
        \includegraphics[width=\linewidth]{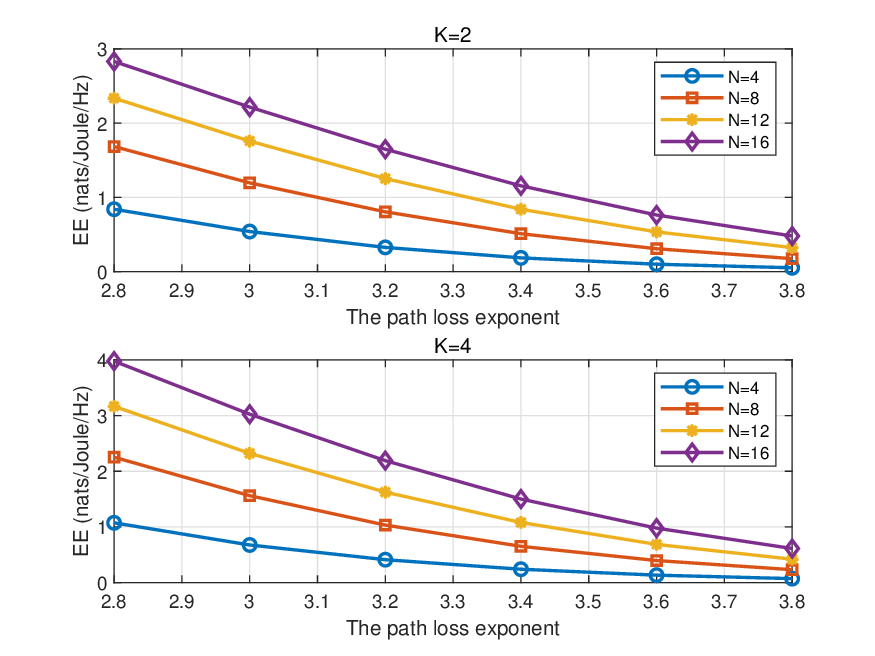}
        \captionof{figure}{EE versus the path-loss exponent.}
        \label{fig.6}
    \end{minipage}
\end{figure}

Fig.~\ref{fig.5} demonstrates the EE of the ISAC system versus the maximum distance between the TRIS transceiver and the users.
The upper and lower subplots correspond to $K=2$ and $K=4$, respectively.
In this experiment, 
the maximum distance is varied from $50$ to $200$~m.
It is observed that the EE decreases monotonically with the maximum distance for all considered $N$,
since a larger transmission distance causes more severe path loss.
In particular, 
the EE drop is more pronounced when the maximum distance increases from $50$ to $100$~m, 
while the curves gradually flatten for larger distances.
Moreover, for a fixed maximum distance, increasing $N$ consistently improves the EE in both subplots.
Finally, 
for the same $N$ and maximum distance, the case with $K=4$ achieves a higher EE than that with $K=2$.
This is because, 
in the examined user-load regime, 
increasing the number of users provides a larger sum-rate gain, 
while the total power consumption does not increase proportionally.

Fig.~\ref{fig.6} shows the achieved EE versus the path-loss exponent of the TRIS transceiver-user link for different system configurations.
Specifically, the path-loss exponent is varied from $2.8$ to $3.8$ for two user settings and for different TRIS sizes.
As observed from Fig.~\ref{fig.6}, 
the EE decreases monotonically with the path-loss exponent due to the more severe large-scale attenuation.
Moreover, for a fixed path-loss exponent, increasing $N$ consistently improves the EE for both values of $K$, highlighting the benefit of employing more TRIS elements.
In addition, for the same system setting, 
the case with $K=4$ achieves a higher EE than that with $K=2$.
Finally, the absolute EE gap between adjacent $N$ values generally shrinks as the path-loss exponent increases, 
since all curves are compressed toward lower EE values under stronger attenuation.

\subsection{Imperfect CSI Case}

In this subsection,
the simulation results for the imperfect CSI case are presented.
Specifically,
the noise power at the receivers and the path-loss exponent of the TRIS-user links are set to
$-70$~dBm and $2.2$, respectively.

\begin{figure}[t]
    \centering
    \begin{minipage}[t]{0.45\textwidth}
        \centering
        \includegraphics[width=\linewidth]{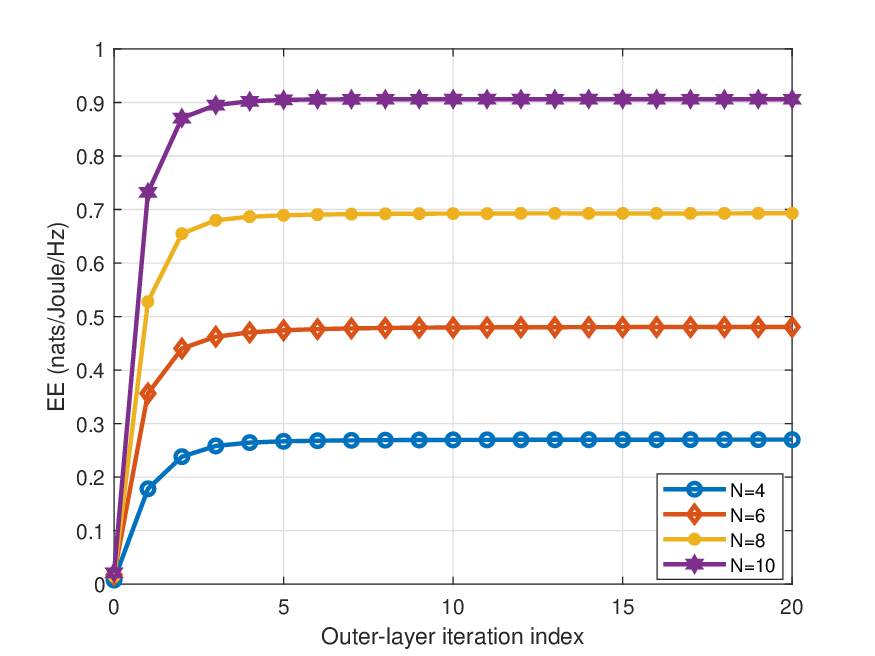}
        \captionof{figure}{{Convergence of Algorithm 2.}}
        \label{fig.7}
    \end{minipage}
    \hfill
    \begin{minipage}[t]{0.45\textwidth}
        \centering
        \includegraphics[width=\linewidth]{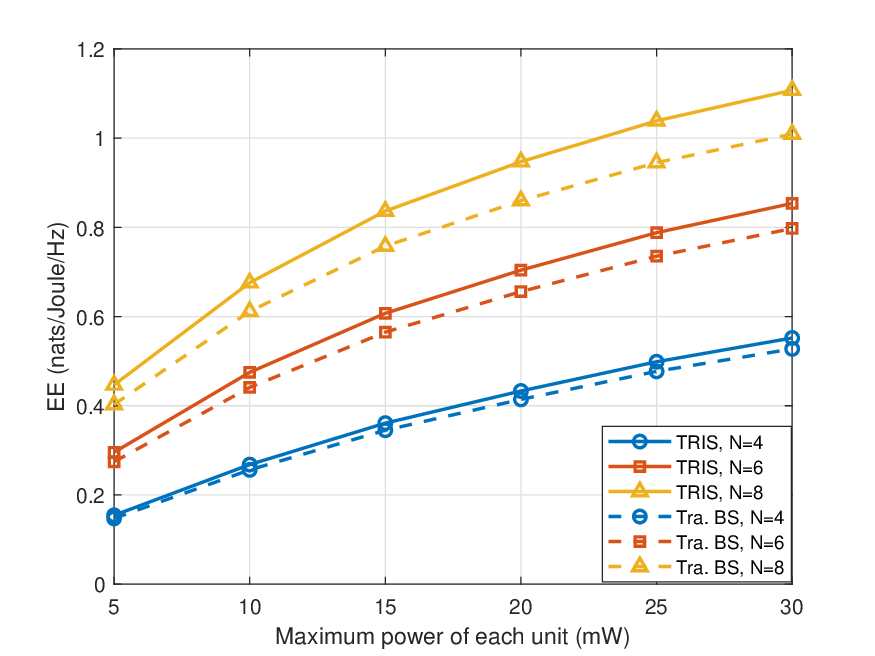}
        \captionof{figure}{{EE versus the maximum power of each TRIS unit.}}
        \label{fig.8}
    \end{minipage}
\end{figure}

\begin{table}[t]
\centering
\caption{{Execution Time of Algorithm~\ref{alg:2} (s)}}
\label{tab:runtime_alg_2}
\renewcommand{\arraystretch}{1.00}
\begin{tabular}{|c|c|c|c|c|}
\hline
\textbf{Algorithm}           & $N=4$  & $N=6$   & $N=8$  & $N=10$   \\
\hline
Algorithm~\ref{alg:2}        & 206.0 & 339.4    & 633.9  & 1050.1 \\
\hline
\end{tabular}
\end{table}

{
First,
we present the convergence behavior of the proposed algorithm when applied to solve the worst-case problem (P1).
As shown in Fig. \ref{fig.7},
we evaluate the convergence performance by varying the number of TRIS elements, i.e., $N=4,6,8,10$.
Notably,
the EE increases monotonically and converges within 10 iterations.
All curves rise sharply within the first 5 iterations and then saturate.
Moreover, 
in all cases, 
a larger number of TRIS units yields higher EE.
Furthermore,
Table~\ref{tab:runtime_alg_2} shows the average execution time of Algorithm~\ref{alg:2} for different numbers of TRIS units.
}

{
In Fig. \ref{fig.8}, 
we investigate the impact of the maximum transmit power of each transmit unit on the EE under different system settings in the imperfect CSI case.
As expected, 
the EE increases with the maximum transmit power for all considered schemes.
In addition, 
increasing the number of transmit units from $N=4$ to $N=8$ consistently improves the EE over the entire power range, 
owing to the enhanced spatial beamforming degrees of freedom.
Furthermore, 
the proposed ``TRIS'' scheme achieves a higher EE than the ``Tra. BS'' scheme under all considered settings.
This is because the TRIS transceiver requires only a single RF chain, 
whereas the fully-active multi-antenna transceiver employs $N$ RF chains, 
which leads to higher circuit power consumption.
Moreover, 
the EE gap between the ``TRIS'' and ``Tra. BS'' schemes tends to increase as the maximum transmit power increases.
}

\begin{figure}[t]
    \centering
    \begin{minipage}[t]{0.45\textwidth}
        \centering
        \includegraphics[width=\linewidth]{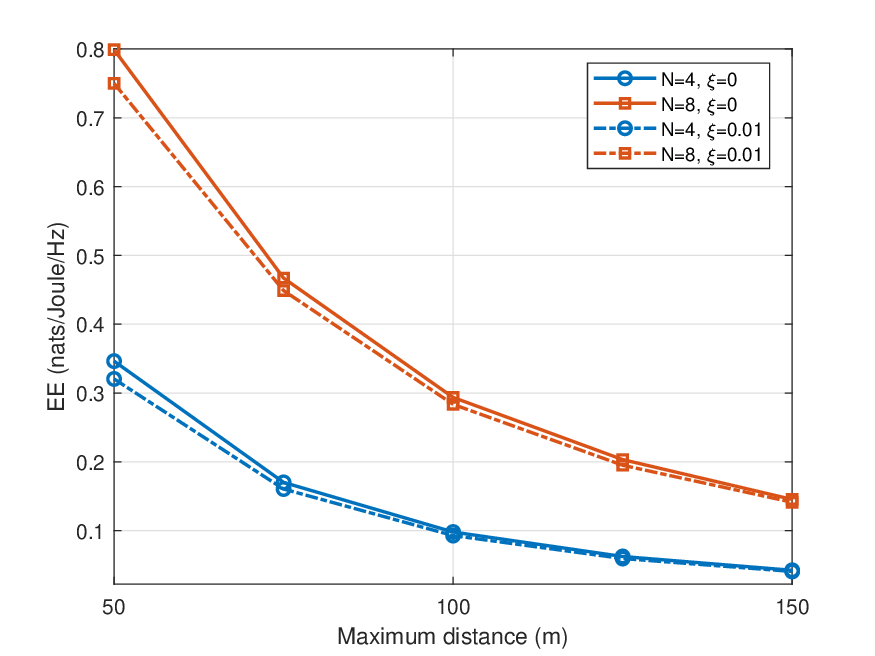}
        \captionof{figure}{EE versus the maximum distance between TRIS transceiver and users.}
        \label{fig.9}
    \end{minipage}
    \hfill
    \begin{minipage}[t]{0.45\textwidth}
        \centering
        \includegraphics[width=\linewidth]{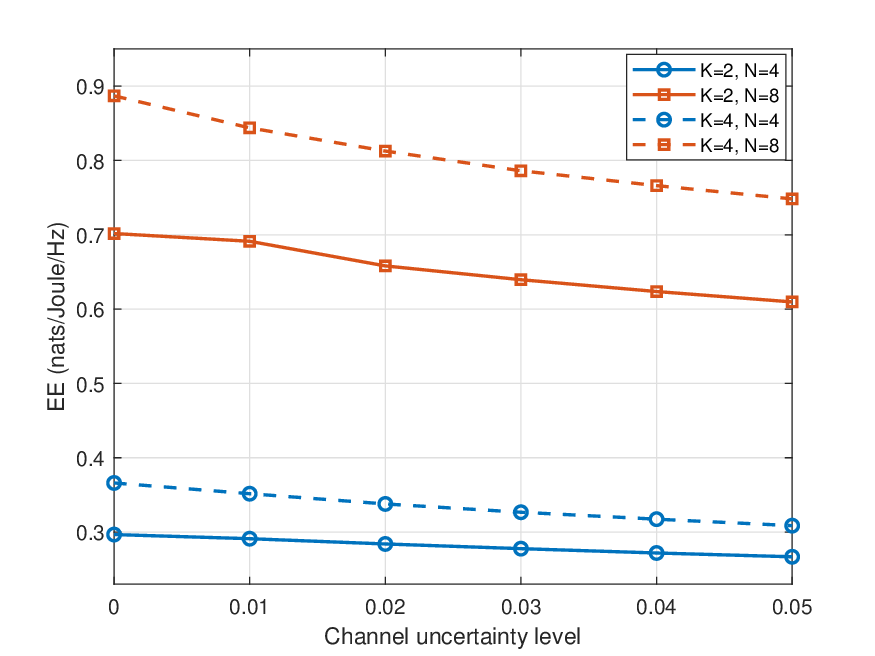}
        \captionof{figure}{EE versus the channel uncertainty level.}
        \label{fig.10}
    \end{minipage}
\end{figure}

Fig. \ref{fig.9} 
illustrates the EE versus the maximum distance between the TRIS transceiver and the users, with $N=4,8$ and $\xi=0,0.01$. 
It can be observed that 
enlarging the distance consistently reduces the EE for all curves.
Also, the curves with $N=8$ stay above those with $N=4$ across the whole distance range. 
In addition, when $\xi=0$, the EE is slightly higher than that with $\xi=0.01$, 
but this difference becomes very small when the distance is large.
This is because, at large distances, 
severe path loss dominates the system performance, 
rendering the additional degradation caused by CSI uncertainty marginal.

Fig. \ref{fig.10} shows the impact of the channel uncertainty level on the system performance.
As illustrated in Fig. \ref{fig.10},
as the channel uncertainty level increases,
the obtained EE decreases for all tested schemes.
This is mainly due to the fact that a higher channel uncertainty level induces more severe CSI mismatch, 
which reduces the beamforming gain achieved by the TRIS transceiver.

\section{Conclusions}

This paper investigates a novel TRIS transceiver-empowered ISAC system to meet the diverse requirements of next-generation wireless networks.
The system EE is maximized by optimizing the transmit beamforming of the TRIS transceiver for the perfect and imperfect CSI scenarios, respectively.
First,
in the perfect CSI case,
we propose an efficient iterative algorithm to solve the non-convex EE maximization problem by using the FP and MM frameworks.
Furthermore,
due to the channel uncertainty,
the algorithm developed for the perfect CSI case cannot be directly applied to the imperfect CSI scenario.
To handle the corresponding semi-infinite inequality constraints,
we adopt the S-Procedure to transform them into LMIs.
Then, 
with the introduction of slack variables and the application of the MM framework,
we develop a tailored algorithm to solve the EE maximization problem in the imperfect CSI case.
The simulation results demonstrate that the proposed algorithms achieve significant EE improvement.

\appendix
\subsection{Proof of Theorem \ref{theorem_1}}
\normalem
{
Proof:
First,
it is obvious that the feasible domains of 
($\{\gamma_{k}\}, \{\omega_{k}\}, \mathbf{x}, y, \{\mathbf{w}_{c,k}\}, \{\mathbf{w}_{r,n}\} $) 
associated with (P2) and (P3) are identical.
Assume that the solution ($\{\gamma_{k}^{(t)}\}, \{\omega_{k}^{(t)}\}, \mathbf{x}^{(t)}, y^{(t)}, \{\mathbf{w}_{c,k}^{(t)}\}, \{\mathbf{w}_{r,n}^{(t)}\} $) is feasible for problem (P3).
Since the updates of the variables $\{\gamma_{k}\}$ and $ \{\omega_{k}\}$ do not decrease $\mathrm{\ddot{R}}_{k}(\{\mathbf{w}_{c,k}\}, \{\mathbf{w}_{r,n}\}, \gamma_{k}, \omega_{k})$ in (\ref{P3_c_1}),
the variables
($\{\gamma_{k}^{(t+1)}\}, \{\omega_{k}^{(t+1)}\}, \mathbf{x}^{(t)}, y^{(t)}, \{\mathbf{w}_{c,k}^{(t)}\}, \{\mathbf{w}_{r,n}^{(t)}\} $) 
remain feasible for problem (P3).
Hence,
the problem (P5)  has a non-empty feasible domain.
By solving problem (P5),
the solution
($\{\gamma_{k}^{(t+1)}\}, \{\omega_{k}^{(t+1)}\}, \mathbf{x}^{(t+1)}, y^{(t+1)}, 
\{\mathbf{w}_{c,k}^{(t+1)}\}, \{\mathbf{w}_{r,n}^{(t+1)}\} $) is obtained and 
it is evidently feasible for problem (P5).
Since the constraint function in (\ref{P5_c_1}) lower-bounds that in (\ref{P3_c_1}) and the constraint function in (\ref{P5_c_2}) lower-bounds that in  (\ref{P3_c_2}), 
($\{\gamma_{k}^{(t+1)}\}, \{\omega_{k}^{(t+1)}\}, \mathbf{x}^{(t+1)}, y^{(t+1)}, 
\{\mathbf{w}_{c,k}^{(t+1)}\}, \{\mathbf{w}_{r,n}^{(t+1)}\} $) is also feasible for problem (P3).

Next,
define 
\begin{align}
&\text{EE}(\{\gamma_{k}\}, \{\omega_{k}\}, \{\mathbf{w}_{c,k}\}, \{\mathbf{w}_{r,n}\})
\triangleq
\frac{
{\sum}_{k=1}^{K}
\mathrm{\ddot{R}}_{k}(\{\mathbf{w}_{c,k}\}, \{\mathbf{w}_{r,n}\}, \gamma_{k}, \omega_{k}) }
{\mathrm{P}_{TRIS}(\{\mathbf{w}_{c,k}\}, \{\mathbf{w}_{r,n}\})}\\
&=\frac{
{\sum}_{k=1}^{K}
\hat{x}^2_{k}(\{\mathbf{w}_{c,k}\}, \{\mathbf{w}_{r,n}\}, \gamma_{k}, \omega_{k}) }
{\hat{y}(\{\mathbf{w}_{c,k}\}, \{\mathbf{w}_{r,n}\})}
\triangleq
\tilde{\text{EE}}(\hat{\mathbf{x}}, \hat{y}),\nonumber
\end{align}
where
\begin{align}
&\hat{x}_{k}(\{\mathbf{w}_{c,k}\}, \{\mathbf{w}_{r,n}\}, \gamma_{k}, \omega_{k}) 
\triangleq 
\sqrt{\mathrm{\ddot{R}}_{k}(\{\mathbf{w}_{c,k}\}, \{\mathbf{w}_{r,n}\}, \gamma_{k}, \omega_{k})} ,\\
&\hat{y}(\{\mathbf{w}_{c,k}\}, \{\mathbf{w}_{r,n}\}) \triangleq \mathrm{P}_{TRIS}(\{\mathbf{w}_{c,k}\}, \{\mathbf{w}_{r,n}\}).\nonumber
\end{align}

Then,
we have
\begin{align}
&\text{EE}(\{\gamma_{k}^{(t+1)}\}, \{\omega_{k}^{(t+1)}\}, \{\mathbf{w}_{c,k}^{(t+1)}\}, \{\mathbf{w}_{r,n}^{(t+1)}\})
\overset{\text{(a)}} =\tilde{\text{EE}}(\hat{\mathbf{x}}^{(t+1)}, \hat{y}^{(t+1)})\\
&\overset{\text{(b)}}\geq 
\tilde{\text{EE}}(\hat{\mathbf{x}}^{(t)}, \hat{y}^{(t)})
+\sum_{k=1}^{K}
\bigg(\frac{2 \hat{{x}}_k^{(t)}}{\hat{y}^{(t)}}
( \hat{{x}}_k^{(t+1)}\! -\! \hat{{x}}_k^{(t)}  )
\!-\! \frac{(\hat{{x}}_k^{(t)})^2}{(\hat{y}^{(t)})^2}
( \mathrm{P}_{TRIS}(\{\mathbf{w}_{c,k}^{(t+1)}\}, \{\mathbf{w}_{r,n}^{(t+1)}\})\! -\! \hat{y}^{(t)} )
 \bigg)\nonumber\\
 &\overset{\text{(c)}} = 
\tilde{\text{EE}}(\hat{\mathbf{x}}^{(t)}, \hat{y}^{(t)})
+\sum_{k=1}^{K}
\bigg(   \frac{2 \hat{x}_k^{(t)}}{\hat{y}^{(t)}}
( {x}_k^{(t+1)} - \hat{x}_k^{(t)}  )
- \frac{(\hat{{x}}_k^{(t)})^2}{(\hat{y}^{(t)})^2}
( y^{(t+1)} - \hat{y}^{(t)} )
 \bigg)\nonumber\\
  &\overset{\text{(d)}} \geq 
\tilde{\text{EE}}(\hat{\mathbf{x}}^{(t)}, \hat{y}^{(t)}) 
+\sum_{k=1}^{K}
\bigg(   \frac{2 \hat{{x}}_k^{(t)}}{\hat{y}^{(t)}}
( {x}_k^{(t)} - \hat{x}_k^{(t)}  )\nonumber  
- \frac{(\hat{x}_k^{(t)})^2}{(\hat{y}^{(t)})^2}
( y^{(t)} - \hat{y}^{(t)} )
 \bigg)\\
&=\tilde{\text{EE}}\big(\hat{\mathbf{x}}^{(t)}, \hat{y}^{(t)}\big)
=\text{EE}\big(\{\gamma_{k}^{(t+1)}\}, \{\omega_{k}^{(t+1)}\}, \{\mathbf{w}_{c,k}^{(t)}\}, \{\mathbf{w}_{r,n}^{(t)}\}\big) \nonumber\\
& \overset{\text{(e)}} \geq  \text{EE}\big(\{\gamma_{k}^{(t)}\}, \{\omega_{k}^{(t)}\}, 
\{\mathbf{w}_{c,k}^{(t)}\}, \{\mathbf{w}_{r,n}^{(t)}\}\big), \nonumber
\end{align}
where
\begin{align}
&\hat{x}_k^{(t+1)}\! \triangleq \!
\hat{x}_k\big(\gamma_{k}^{(t+1)}, \omega_{k}^{(t+1)}, 
\{\mathbf{w}_{c,k}^{(t+1)}\}, \{\mathbf{w}_{r,n}^{(t+1)}\}\big),
 \hat{x}_k^{(t)}\! \triangleq \!
\hat{x}_k\big(\gamma_{k}^{(t+1)}, \omega_{k}^{(t+1)}, 
\{\mathbf{w}_{c,k}^{(t)}\}, \{\mathbf{w}_{r,n}^{(t)}\}\big).
\end{align}

First, the equality (a) follows directly from the prior definitions.
Based on (\ref{EE_MM_1}), the inequality (b) holds.
When problem (P5) is solved, 
constraints (\ref{P5_c_1}) and (\ref{P5_c_1_2}) can be satisfied with equality.
Otherwise, increasing $x_k$ or decreasing $y$ preserves feasibility and does not decrease the objective (\ref{P5_obj}). 
Hence,
$\hat{x}_k^{(t+1)} = x_k^{(t+1)}$
and
$\hat{y}^{(t+1)} = y^{(t+1)}$.
Therefore, equality (c) follows.
The inequality (d) is satisfied since the update of 
($\{\mathbf{w}_{c,k}\}, \{\mathbf{w}_{r,n}\}$) 
obtained by solving problem (P5) yields an increase in the objective value.
Furthermore,
the inequality (e) follows from the fact that the updates of $\{\gamma_{k}\}$
and
$\{\omega_{k}\}$ do not decrease the numerator of the objective (\ref{P2_obj}).
Hence, the value of (\ref{P2_obj}) increases monotonically.
Based on the above discussion, 
the objective value of (P2) is upper-bounded. 
Therefore, the sequence of its objective values converges.
}

\subsection{Detailed derivation for the constraint (\ref{P7_c_2})}
\normalem

{
First,
by using the MM method,
we can obtain a lower bound of 
the term 
$\vert {\mathbf{h}}_{c,k}^H\mathbf{w}_{c,i}\vert^2$ 
in (\ref{P7_c_2}) as follows}
\begin{align}
&\vert {\mathbf{h}}_{c,k}^H\mathbf{w}_{c,i}\vert^2 
\geq
2\text{Re}\{ ({\mathbf{h}}_{c,k}^H\mathbf{w}_{c,i,0})^H {\mathbf{h}}_{c,k}^H\mathbf{w}_{c,i} \}
-
\vert {\mathbf{h}}_{c,k}^H\mathbf{w}_{c,i,0}\vert^2.\label{P7_c_2_1}
\end{align}

In the following, 
we rewrite terms in (\ref{P7_c_2_1}) into a compact form w.r.t. $\Delta\mathbf{h}_{c,k}$.
Specifically, 
by substituting $\mathbf{h}_{c,k} = \hat{\mathbf{h}}_{c,k} + \Delta\mathbf{h}_{c,k}$
into $\vert {\mathbf{h}}_{c,k}^H\mathbf{w}_{c,i,0}\vert^2$,
we can expand it as follows
\begin{align}
&\vert {\mathbf{h}}_{c,k}^H\mathbf{w}_{c,i,0}\vert^2 
=\mathbf{w}_{c,i,0}^H{\mathbf{h}}_{c,k} {\mathbf{h}}_{c,k}^H\mathbf{w}_{c,i,0}\label{P7_c_2_1_1}\\
&= \mathbf{w}_{c,i,0}^H(\hat{\mathbf{h}}_{c,k} + \Delta\mathbf{h}_{c,k})
(\hat{\mathbf{h}}_{c,k} + \Delta\mathbf{h}_{c,k})^H\mathbf{w}_{c,i,0}\nonumber\\
&=\mathbf{w}_{c,i,0}^H\hat{\mathbf{h}}_{c,k} \hat{\mathbf{h}}_{c,k}^H\mathbf{w}_{c,i,0}
+\mathbf{w}_{c,i,0}^H\hat{\mathbf{h}}_{c,k} \Delta{\mathbf{h}}_{c,k}^H\mathbf{w}_{c,i,0}\nonumber\\
&+\mathbf{w}_{c,i,0}^H\Delta{\mathbf{h}}_{c,k} \hat{\mathbf{h}}_{c,k}^H\mathbf{w}_{c,i,0}
+\mathbf{w}_{c,i,0}^H\Delta{\mathbf{h}}_{c,k} \Delta{\mathbf{h}}_{c,k}^H\mathbf{w}_{c,i,0}\nonumber\\
&=\Delta\mathbf{h}_{c,k}^H\mathbf{B}_{1,i}\Delta\mathbf{h}_{c,k} + 
2\text{Re}\{ \mathbf{b}_{1,k,i}^H \Delta\mathbf{h}_{c,k}\} + c_{1,k,i},\nonumber
\end{align}
where
\begin{align}
c_{1,k,i} \triangleq \hat{\mathbf{h}}_{c,k}^H \mathbf{w}_{c,i,0}\mathbf{w}_{c,i,0}^H  \hat{\mathbf{h}}_{c,k},
\mathbf{b}_{1,k,i} \triangleq \mathbf{w}_{c,i,0}\mathbf{w}_{c,i,0}^H  \hat{\mathbf{h}}_{c,k},
\mathbf{B}_{1,i} \triangleq \mathbf{w}_{c,i,0}\mathbf{w}_{c,i,0}^H.
\end{align}

Furthermore,
$2\text{Re}\{ ({\mathbf{h}}_{c,k}^H\mathbf{w}_{c,i,0})^H {\mathbf{h}}_{c,k}^H\mathbf{w}_{c,i} \}$
can be compactly rewritten as
\begin{align}
&2\text{Re}\{ ({\mathbf{h}}_{c,k}^H\mathbf{w}_{c,i,0})^H {\mathbf{h}}_{c,k}^H\mathbf{w}_{c,i} \}\label{P7_c_2_1_2}\\
&=2\text{Re}\{ ((\hat{\mathbf{h}}_{c,k} + \Delta\mathbf{h}_{c,k})^H\mathbf{w}_{c,i,0})^H (\hat{\mathbf{h}}_{c,k} + \Delta\mathbf{h}_{c,k})^H\mathbf{w}_{c,i} \}\nonumber\\
&= 2\text{Re}\{
\mathbf{w}_{c,i,0}^H\hat{\mathbf{h}}_{c,k} \hat{\mathbf{h}}_{c,k}^H\mathbf{w}_{c,i}
+\mathbf{w}_{c,i,0}^H\hat{\mathbf{h}}_{c,k} \Delta{\mathbf{h}}_{c,k}^H\mathbf{w}_{c,i}\nonumber\\
&+\mathbf{w}_{c,i,0}^H\Delta{\mathbf{h}}_{c,k} \hat{\mathbf{h}}_{c,k}^H\mathbf{w}_{c,i}
+\mathbf{w}_{c,i,0}^H\Delta{\mathbf{h}}_{c,k} \Delta{\mathbf{h}}_{c,k}^H\mathbf{w}_{c,i}
\}
\nonumber\\
&=\Delta\mathbf{h}_{c,k}^H(\mathbf{B}_{2,i} +\mathbf{B}_{2,i}^H )\Delta\mathbf{h}_{c,k}
+ 
2\text{Re}\{ (\mathbf{b}_{2,k,i}+\mathbf{b}_{3,k,i})^H \Delta\mathbf{h}_{c,k}\} + c_{2,k,i},
\nonumber
\end{align}
where
\begin{align}
&\mathbf{b}_{3,k,i} \triangleq \mathbf{w}_{c,i,0}\mathbf{w}_{c,i}^H  \hat{\mathbf{h}}_{c,k},
\mathbf{B}_{2,i} \triangleq \mathbf{w}_{c,i}\mathbf{w}_{c,i,0}^H,
c_{2,k,i} \triangleq 2\text{Re}\{ \hat{\mathbf{h}}_{c,k}^H \mathbf{w}_{c,i}\mathbf{w}_{c,i,0}^H  \hat{\mathbf{h}}_{c,k} \},\\
&\mathbf{b}_{2,k,i} \triangleq \mathbf{w}_{c,i}\mathbf{w}_{c,i,0}^H  \hat{\mathbf{h}}_{c,k}.\nonumber
\end{align}

Based on (\ref{P7_c_2_1_1}) and (\ref{P7_c_2_1_2}),
the right-hand side of (\ref{P7_c_2_1}) can be rewritten as
\begin{align}
&2\text{Re}\{ ({\mathbf{h}}_{c,k}^H\mathbf{w}_{c,i,0})^H {\mathbf{h}}_{c,k}^H\mathbf{w}_{c,i} \}
-
\vert {\mathbf{h}}_{c,k}^H\mathbf{w}_{c,i,0}\vert^2\\
&\Leftrightarrow
\Delta\mathbf{h}_{c,k}^H\mathbf{B}_{3,i} \Delta\mathbf{h}_{c,k} + 
2\text{Re}\{ \mathbf{b}_{4,k,i}^H \Delta\mathbf{h}_{c,k}\} + c_{3,k,i},\nonumber
\end{align}
where
\begin{align}
&c_{3,k,i} \triangleq c_{2,k,i} - c_{1,k,i},
\mathbf{b}_{4,k,i} \triangleq \mathbf{b}_{3,k,i} + \mathbf{b}_{2,k,i} - \mathbf{b}_{1,k,i},
\mathbf{B}_{3,i} \triangleq \mathbf{B}_{2,i} + \mathbf{B}_{2,i}^H - \mathbf{B}_{1,i}.
\end{align}

Using the MM method,
the term 
$\vert {\mathbf{h}}_{c,k}^H\mathbf{w}_{r,n}\vert^2$
in 
(\ref{P7_c_2})
can be lower-bounded as follows
\begin{align}
&\vert {\mathbf{h}}_{c,k}^H  \mathbf{w}_{r,n}\vert^2
\geq
2\text{Re}\{ ({\mathbf{h}}_{c,k}^H\mathbf{w}_{r,n,0})^H{\mathbf{h}}_{c,k}^H\mathbf{w}_{r,n}\}
-
\vert {\mathbf{h}}_{c,k}^H\mathbf{w}_{r,n,0}\vert^2\\
&=\Delta\mathbf{h}_{c,k}^H\mathbf{B}_{7,n}\Delta\mathbf{h}_{c,k} + 
2\text{Re}\{ \mathbf{b}_{9,k,n}^H \Delta\mathbf{h}_{c,k}\} + c_{7,k,n},\nonumber
\end{align}
where the newly introduced coefficients are defined in (\ref{Coef_P7_c1_SCA_1}).
\begin{figure*}
\begin{small}
\begin{align}
&c_{5,k,n}\!\! \triangleq\!\! \hat{\mathbf{h}}_{c,k}^H \mathbf{w}_{r,n,0}\mathbf{w}_{r,n,0}^H  \hat{\mathbf{h}}_{c,k},
\mathbf{b}_{6,k,n}\!\! \triangleq\!\! \mathbf{w}_{r,n,0}\mathbf{w}_{r,n,0}^H  \hat{\mathbf{h}}_{c,k},
\mathbf{B}_{5,n}\! \triangleq\! \mathbf{w}_{r,n,0}\mathbf{w}_{r,n,0}^H,
c_{6,k,n}\! \triangleq\! 2\text{Re}\{ \hat{\mathbf{h}}_{c,k}^H \mathbf{w}_{r,n}\mathbf{w}_{r,n,0}^H\hat{\mathbf{h}}_{c,k} \},\nonumber\\
&\mathbf{b}_{7,k,n} \triangleq \mathbf{w}_{r,n}\mathbf{w}_{r,n,0}^H  \hat{\mathbf{h}}_{c,k},
\mathbf{b}_{8,k,n} \triangleq \mathbf{w}_{r,n,0}\mathbf{w}_{r,n}^H  \hat{\mathbf{h}}_{c,k},
\mathbf{B}_{6,n} \triangleq \mathbf{w}_{r,n}\mathbf{w}_{r,n,0}^H,
c_{7,k,n} \triangleq c_{6,k,n} - c_{5,k,n},\nonumber\\
&\mathbf{b}_{9,k,n} \triangleq \mathbf{b}_{7,k,n} + \mathbf{b}_{8,k,n} - \mathbf{b}_{6,k,n},
\mathbf{B}_{7,n} \triangleq \mathbf{B}_{6,n} + \mathbf{B}_{6,n}^H - \mathbf{B}_{5,n}.\label{Coef_P7_c1_SCA_1}
\end{align}
\end{small}
\boldsymbol{\hrule}
\end{figure*}

Thus,
based on the above transformations,
the constraint (\ref{P7_c_2}) can be rewritten as
\begin{align}
\Delta\mathbf{h}_{c,k}^H\mathbf{B}_{9}\Delta\mathbf{h}_{c,k}
\! +\! 
2\text{Re}\{ \mathbf{b}_{11,k}^H \Delta\mathbf{h}_{c,k}\}\! +\! c_{9,k}+\sigma_{c,k}^2
\!\geq\! \mu_k, \label{P7_c_2_T1}
\end{align}
where
\begin{align}
&\mathbf{B}_{9} \triangleq {\sum}_{i=1}^{K}\mathbf{B}_{3,i} + {\sum}_{n=1}^{N}\mathbf{B}_{7,n},
\mathbf{b}_{11,k} \triangleq {\sum}_{i=1}^{K}\mathbf{b}_{4,k,i} + {\sum}_{n=1}^{N}\mathbf{b}_{9,k,n},\\
&c_{9,k} \triangleq {\sum}_{i=1}^{K}c_{3,k,i} + {\sum}_{n=1}^{N}c_{7,k,n}.  \nonumber
\end{align}

{
It is worth noting that (\ref{P7_c_2_T1}) is not an exact reformulation of (\ref{P7_c_2}), 
but provides a conservative sufficient condition for it. 
According to the property of the MM method, 
if 
(\ref{P7_c_2_T1}) is satisfied for all 
$\Vert\Delta\mathbf{h}_{c,k}\Vert_2 \leq \xi_{c,k}$, 
the original constraint (\ref{P7_c_2}) is also guaranteed to hold.}

Note that (\ref{P7_c_2_T1}) is still a semi-infinite inequality constraint due to the CSI uncertainty.
By leveraging the S-Procedure,
(\ref{P7_c_2_T1}) 
can be equivalently transformed into the following LMIs as
\begin{align}
&\Delta\mathbf{h}_{c,k}^H\mathbf{B}_{9}\Delta\mathbf{h}_{c,k}
\! +\! 
2\text{Re}\{ \mathbf{b}_{11,k}^H \Delta\mathbf{h}_{c,k}\}\! +\! c_{9,k}+\sigma_{c,k}^2
\!\geq\! \mu_k \\
&\Rightarrow
\left[
\begin{array}{cc}
\varpi_{1,k} \mathbf{I} + \mathbf{B}_{9} & \mathbf{b}_{11,k} \\
\mathbf{b}_{11,k}^H  & c_{9,k}+\sigma_{c,k}^2 - \mu_k - \varpi_{1,k}\xi_{c,k}^2
\end{array}
\right] \succeq \mathbf{0},\nonumber
\end{align}
where
$ \boldsymbol{\varpi}_1=[\varpi_{1,1},\cdots,\varpi_{1,K}]^T \geq 0   $.

\subsection{Detailed derivation for the constraint (\ref{P7_c_3})}
\normalem

To circumvent the CSI uncertainty in constraint (\ref{P7_c_3}),
by adopting Schur's complement Lemma \cite{ref_Convex Optimization},
the inequality constraint (\ref{P7_c_3}) can be recast into matrix form as follows
\begin{align}
&{\sum}_{i\neq k}^{K}\vert {\mathbf{h}}_{c,k}^H\mathbf{w}_{c,i}\vert^2
+{\sum}_{n=1}^{N} \vert {\mathbf{h}}_{c,k}^H\mathbf{w}_{r,n}\vert^2
+\sigma_{c,k}^2\leq \nu_k \label{P7_c_3_T1}\\
&\Rightarrow
\left[
\begin{array}{cc}
\nu_k - \sigma_{c,k}^2 & \mathbf{h}_{c,k}^H \mathbf{W}_{cr,-k}\\
\mathbf{W}_{cr,-k}^H\mathbf{h}_{c,k}  &  \mathbf{I}
\end{array}
\right] \succeq \mathbf{0},\nonumber
\end{align}
where
\begin{align}
\mathbf{W}_{cr,-k} \triangleq 
&[ \mathbf{w}_{c,1},\cdots ,\mathbf{w}_{c,k-1},\mathbf{w}_{c,k+1},  \cdots, \mathbf{w}_{c,K},\\
& \mathbf{w}_{r,1}, \cdots, \mathbf{w}_{r,N} ] \in \mathbb{C}^{N\times(K-1+N)} .\nonumber
\end{align}

By substituting $\mathbf{h}_{c,k} = \hat{\mathbf{h}}_{c,k} + \Delta\mathbf{h}_{c,k}$ into (\ref{P7_c_3_T1}),
we can obtain
\begin{align}
&\left[
\begin{array}{cc}
\nu_k - \sigma_{c,k}^2 & (\hat{\mathbf{h}}_{c,k} + \Delta\mathbf{h}_{c,k})^H \mathbf{W}_{cr,-k}\\
\mathbf{W}_{cr,-k}^H(\hat{\mathbf{h}}_{c,k} + \Delta\mathbf{h}_{c,k})  &  \mathbf{I}
\end{array}
\right]\succeq \mathbf{0}\label{P7_c_3_T2}\\
&\Leftrightarrow
\left[
\begin{array}{cc}
\nu_k - \sigma_{c,k}^2 & \hat{\mathbf{h}}_{c,k}^H \mathbf{W}_{cr,-k}\\
\mathbf{W}_{cr,-k}^H\hat{\mathbf{h}}_{c,k}  &  \mathbf{I}
\end{array}
\right]
-
\left[\!\!\!
\begin{array}{cc}
1\\
\mathbf{0}
\end{array}\!\!\!
\right]
\Delta \mathbf{h}_{c,k}^H
\left[\!\!\!
\begin{array}{cc}
\mathbf{0},
\mathbf{W}_{cr,-k}
\end{array}\!\!\!
\right]
-
\left[\!\!\!
\begin{array}{cc}
\mathbf{0}\\
\mathbf{W}_{cr,-k}
\end{array}\!\!\!
\right]
\Delta\mathbf{h}_{c,k}
\left[\!\!\!
\begin{array}{cc}
1,
\mathbf{0}
\end{array}\!\!\!
\right]
\succeq \mathbf{0}
\nonumber
\\
&
\Leftrightarrow
\left[
\begin{array}{cc}
\nu_k - \sigma_{c,k}^2 & \hat{\mathbf{h}}_{c,k}^H \mathbf{W}_{cr,-k}\\
\mathbf{W}_{cr,-k}^H\hat{\mathbf{h}}_{c,k}  &  \mathbf{I}
\end{array}
\right]
\succeq 
\left[\!\!\!
\begin{array}{cc}
1\\
\mathbf{0}
\end{array}\!\!\!
\right]
\Delta \mathbf{h}_{c,k}^H
\left[\!\!\!
\begin{array}{cc}
\mathbf{0},
-\mathbf{W}_{cr,-k}
\end{array}\!\!\!
\right]
+
\left[\!\!\!
\begin{array}{cc}
\mathbf{0}\\
-\mathbf{W}_{cr,-k}
\end{array}\!\!\!
\right]
\Delta\mathbf{h}_{c,k}
\left[\!\!\!
\begin{array}{cc}
1,
\mathbf{0}
\end{array}\!\!\!
\right]
.\nonumber
\end{align}

By using the above sign-definiteness lemma,
the equivalent LMIs of (\ref{P7_c_3_T2}) can be written as
\begin{align}
&
\left[
\begin{array}{ccc}
\nu_k - \sigma_{c,k}^2-\varpi_{2,k} & \hat{\mathbf{h}}_{c,k}^H \mathbf{W}_{cr,-k} & \mathbf{0}\\
\mathbf{W}_{cr,-k}^H\hat{\mathbf{h}}_{c,k}  &  \mathbf{I}  & \xi_{c,k}\mathbf{W}_{cr,-k}^H\\
\mathbf{0}  &  \xi_{c,k}\mathbf{W}_{cr,-k}  & \varpi_{2,k}\mathbf{I}
\end{array}
\right] 
\succeq 
\mathbf{0},
\end{align}
where
$ \boldsymbol{\varpi}_2=[\varpi_{2,1},\cdots,\varpi_{2,K}]^T \geq 0   $.

\subsection{Detailed derivation for the constraint (\ref{P7_c_6})}
\normalem

First,
we convexify the terms 
$\vert{\mathbf{h}}_{r}^H \mathbf{w}_{r,i}\vert^2 $
and 
$\vert{\mathbf{h}}_{r}^H \mathbf{w}_{c,k}\vert^2$ 
in (\ref{P7_c_6})
via invoking the MM method as follows
\begin{align}
&\vert{\mathbf{h}}_{r}^H \mathbf{w}_{r,i}\vert^2
\geq
2\text{Re}\{(\mathbf{h}_{r}^H \mathbf{w}_{r,i,0})^H\mathbf{h}_{r}^H \mathbf{w}_{r,i}\}
-\vert\mathbf{h}_{r}^H \mathbf{w}_{r,i,0}\vert^2
\\
&=\Delta\mathbf{h}_{r}^H\mathbf{B}_{12,i}\Delta\mathbf{h}_{r} + 
2\text{Re}\{ \mathbf{b}_{15,i}^H \Delta\mathbf{h}_{r}\} + c_{12,i},\nonumber
\\
&\vert{\mathbf{h}}_{r}^H \mathbf{w}_{c,k}\vert^2
\geq
2\text{Re}\{(\mathbf{h}_{r}^H \mathbf{w}_{c,k,0})^H\mathbf{h}_{r}^H \mathbf{w}_{c,k}\}
-\vert\mathbf{h}_{r}^H \mathbf{w}_{c,k,0}\vert^2
\\
&=\Delta\mathbf{h}_{r}^H\mathbf{B}_{16,k}\Delta\mathbf{h}_{r} + 
2\text{Re}\{ \mathbf{b}_{20,k}^H \Delta\mathbf{h}_{r}\} + c_{16,k},\nonumber
\end{align}
respectively,
where the newly introduced notations above are defined in (\ref{Coef_P7_c2_SCA_1}).
\begin{figure*}
\begin{small}
\begin{align}
&c_{10,n}\!\triangleq\! 
\hat{\mathbf{h}}_{r}^H \mathbf{w}_{r,n,0}\mathbf{w}_{r,n,0}^H  \hat{\mathbf{h}}_{r},
\mathbf{b}_{12,n}\triangleq 
\mathbf{w}_{r,n,0}\mathbf{w}_{r,n,0}^H  \hat{\mathbf{h}}_{r},
\mathbf{B}_{10,n}\triangleq 
\mathbf{w}_{r,n,0}\mathbf{w}_{r,n,0}^H,
c_{11,n} \triangleq 2\text{Re}\{ \hat{\mathbf{h}}_{r}^H \mathbf{w}_{r,n}\mathbf{w}_{r,n,0}^H\hat{\mathbf{h}}_{r} \},
\label{Coef_P7_c2_SCA_1}\\
&
\mathbf{b}_{13,n}\!\triangleq \!
\mathbf{w}_{r,n,0}\mathbf{w}_{r,n}^H  \hat{\mathbf{h}}_{r},
\mathbf{b}_{14,n}\triangleq 
\mathbf{w}_{r,n}\mathbf{w}_{r,n,0}^H  \hat{\mathbf{h}}_{r},
\mathbf{B}_{11,n}\!\triangleq \!
\mathbf{w}_{r,n}\mathbf{w}_{r,n,0}^H,
c_{12,n}\triangleq  c_{11,n}\! -\! c_{10,n},
\mathbf{b}_{15,n}\triangleq  \mathbf{b}_{14,n} + \mathbf{b}_{13,n} - \mathbf{b}_{12,n},\nonumber\\
&\mathbf{B}_{12,n}\triangleq 
\mathbf{B}_{11,n} + \mathbf{B}_{11,n}^H - \mathbf{B}_{10,n},
c_{14,k}\triangleq 
\hat{\mathbf{h}}_{r}^H \mathbf{w}_{c,k,0}\mathbf{w}_{c,k,0}^H  \hat{\mathbf{h}}_{r},
\mathbf{b}_{17,k}\triangleq 
\mathbf{w}_{c,k,0}\mathbf{w}_{c,k,0}^H  \hat{\mathbf{h}}_{r},
\mathbf{B}_{14,k}\triangleq 
\mathbf{w}_{c,k,0}\mathbf{w}_{c,k,0}^H,
\nonumber\\
&c_{15,k} \triangleq 2\text{Re}\{ \hat{\mathbf{h}}_{r}^H \mathbf{w}_{c,k}\mathbf{w}_{c,k,0}^H\hat{\mathbf{h}}_{r} \},
\mathbf{b}_{18,k}\triangleq 
\mathbf{w}_{c,k,0}\mathbf{w}_{c,k}^H  \hat{\mathbf{h}}_{r},
\mathbf{b}_{19,k}\triangleq 
\mathbf{w}_{c,k}\mathbf{w}_{c,k,0}^H  \hat{\mathbf{h}}_{r},
\mathbf{B}_{15,k}\triangleq 
\mathbf{w}_{c,k}\mathbf{w}_{c,k,0}^H,\nonumber\\
&c_{16,k}\triangleq  c_{15,k} - c_{14,k},
\mathbf{b}_{20,k}\triangleq  \mathbf{b}_{18,k} + \mathbf{b}_{19,k} - \mathbf{b}_{17,k},
\mathbf{B}_{16,k}\triangleq 
\mathbf{B}_{15,k} + \mathbf{B}_{15,k}^H - \mathbf{B}_{14,k}.\nonumber
\end{align}
\end{small}
\boldsymbol{\hrule}
\end{figure*}

Therefore,
the constraint (\ref{P7_c_6}) can be reformulated as follows
\begin{align}
\Delta\mathbf{h}_{r}^H\mathbf{B}_{18}\Delta\mathbf{h}_{r} + 
2\text{Re}\{ \mathbf{b}_{22}^H \Delta\mathbf{h}_{r}\} + c_{18}
\geq P_r, \label{P7_c_6_T_1}
\end{align}
where
\begin{align}
&\mathbf{B}_{18} \triangleq {\sum}_{k=1}^{K}\mathbf{B}_{16,k} + {\sum}_{n=1}^{N}\mathbf{B}_{12,n},
\mathbf{b}_{22} \triangleq {\sum}_{k=1}^{K}\mathbf{b}_{20,k} + {\sum}_{n=1}^{N}\mathbf{b}_{15,n},\\
&c_{18} \triangleq {\sum}_{k=1}^{K}c_{16,k} + {\sum}_{n=1}^{N}c_{12,n}. \nonumber
\end{align}

By applying the S-Procedure again,
the following LMI of (\ref{P7_c_6_T_1}) can be given as
\begin{align}
&\Delta\mathbf{h}_{r}^H\mathbf{B}_{18}\Delta\mathbf{h}_{r} + 
2\text{Re}\{ \mathbf{b}_{22}^H \Delta\mathbf{h}_{r}\} + c_{18}
\geq P_r\\
&\Rightarrow
\left[
\begin{array}{cc}
\varpi_{3} \mathbf{I} + \mathbf{B}_{18} & \mathbf{b}_{22} \\
\mathbf{b}_{22}^H  & c_{18}-P_r - \varpi_{3}\xi_{r}^2
\end{array}
\right] \succeq \mathbf{0},\nonumber
\end{align}
where
$ \varpi_3 \geq 0 $ is a slack variable.



\begin{thebibliography}{99}




\bibitem{ref_ISAC_1}
R. Liu, L. Zhang, Y. R. Li, and M. D. Renzo, 
``The ITU vision and  framework for 6G: Scenarios, capabilities, and enablers,'' 
\emph{IEEE Veh. Technol. Mag.}, 
vol. 20, no. 2, pp. 114$-$122, Jun. 2025.


\bibitem{ref_ISAC_2}
F. Liu \emph{et al.}, 
``Integrated sensing and communications: Toward dual-functional wireless networks for 6G and beyond,'' 
\emph{IEEE J. Sel. Areas Commun.}, 
vol. 40, no. 6, pp. 1728$-$1767, Jun. 2022.



\bibitem{ref_ISAC_3}
Z. Zhang, W. Chen, Q. Wu, Z. Li, X. Zhu, and J. Yuan, 
``Intelligent omni-surfaces assisted integrated multi-target sensing and multi-user MIMO communications,'' 
\emph{IEEE Trans. Commun.}, 
vol. 72, no. 8, pp. 4591$-$4606, Aug. 2024.




\bibitem{ref_ISAC_4}
Y. Guo \emph{et al.}, 
``Movable antenna enhanced networked integrated sensing and communication system,'' 
\emph{IEEE Trans. Wireless Commun.}, 
vol. 25, pp. 5555$-$5572, 2026.






\bibitem{ref_ISAC_5}
Y. Guo, W. Chen, Q. Wu, Y. Liu, and Q. Wu, 
``Cram\'{e}r-Rao bound optimization for fluid antenna-empowered integrated sensing and uplink communication system,'' 
\emph{IEEE Trans. Commun.}, 
vol. 74, pp. 3631$-$3645, 2026.



\bibitem{ref_RIS_1}
Q. Wu \emph{et al.}, 
``Intelligent surfaces empowered wireless network: Recent advances and the road to 6G,''
\emph{Proc. IEEE}, 
vol. 112, no. 7, pp. 724$-$763, Jul. 2024.


\bibitem{ref_RIS_2}
Q. Wu, S. Zhang, B. Zheng, C. You, and R. Zhang, 
``Intelligent reflecting surface-aided wireless communications: A tutorial,'' 
\emph{IEEE Trans. Commun.}, 
vol. 69, no. 5, pp. 3313$-$3351, May 2021.

\bibitem{ref_RIS_3}
Y. Guo, Y. Liu, Q. Wu, Q. Shi, and Y. Zhao, 
``Enhanced secure communication via novel double-faced active RIS,'' 
\emph{IEEE Trans. Commun.}, 
vol. 71, no. 6, pp. 3497$-$3512, Jun. 2023.


\bibitem{ref_RIS_ISAC_0}
Y. Guo, Y. Liu, Q. Wu, X. Li, and Q. Shi, 
``Joint beamforming and power allocation for RIS aided full-duplex integrated sensing and uplink communication system,'' 
\emph{IEEE Trans. Wireless Commun.}, 
vol. 23, no. 5, pp. 4627$-$4642, May 2024.

\bibitem{ref_RIS_ISAC_0_1}
Z. Guang, Y. Liu, Q. Wu, Y. -F. Liu, and Q. Shi, 
``Communication aided sensing for RIS assisted MU-MIMO system: CRB optimization with guaranteed ergodic rate,'' 
\emph{IEEE Trans. Signal Process.}, 
vol. 73, pp. 4208$-$4225, 2025.




\bibitem{ref_RIS_ISAC_1}
Q. Wu \emph{et al.},
``Intelligent reflecting surfaces for integrated sensing and communications: A survey,''
Nov. 2025.
[Online]. 
Available: https://arxiv.org/abs/2511.10990


\bibitem{ref_TRIS_1}
Z. Li \emph{et al.}, 
``Transmissive reconfigurable intelligent surface-enabled transceiver systems: Architecture, design issues, and opportunities,''
\emph{IEEE Veh. Technol. Mag.}, 
vol. 19, no. 4, pp. 44$-$53, Dec. 2024.



\bibitem{ref_RIS_transceiver_1}
W. Tang \emph{et al.}, 
``MIMO transmission through reconfigurable intelligent surface: System design, analysis, and implementation,'' 
\emph{IEEE J. Sel. Areas Commun.}, 
vol. 38, no. 11, pp. 2683$-$2699, Nov. 2020.





\bibitem{ref_TRIS_app_1}
Z. Li \emph{et al.}, 
``Toward TMA-based transmissive RIS transceiver enabled downlink communication networks: A consensus-ADMM approach,''
\emph{IEEE Trans. Commun.}, 
vol. 73, no. 4, pp. 2832$-$2846, Apr. 2025.




\bibitem{ref_TRIS_app_1_1}
Y. Guo, W. Chen, Y. Zhu, Z. Li, Q. Wu, and K. Wang,
``Beamforming for transmissive RIS transceiver enabled simultaneous wireless information and power transfer systems,'' 
Nov. 2025.
[Online]. 
Available: https://arxiv.org/abs/2511.11985


\bibitem{ref_TRIS_app_1_2}
Y. Guo, W. Chen, X. Bai, C. He, and Q. Wu, 
``Resource allocation for transmissive RIS transceiver-enabled SWIPT systems,'' 
\emph{IEEE Trans. Veh. Technol.}, 
early access,
June 04, 2026,
doi: 10.1109/TVT.2026.3700100.









\bibitem{ref_TRIS_app_2}
Z. Li, W. Chen, Z. Zhang, Q. Wu, H. Cao, and J. Li, 
``Robust sum-rate maximization in transmissive RMS transceiver-enabled SWIPT networks,'' 
\emph{IEEE Internet Things J.}, 
vol. 10, no. 8, pp. 7259$-$7271, Apr. 2023.

\bibitem{ref_TRIS_app_3}
Z. Li, W. Chen, Z. Liu, H. Tang, and J. Lu, 
``Joint communication and computation design in transmissive RMS transceiver enabled multi-tier computing networks,'' 
\emph{IEEE J. Sel. Areas Commun.}, 
vol. 41, no. 2, pp. 334$-$348, Feb. 2023.

\bibitem{ref_TRIS_app_4}
A. Huang, X. Mu, L. Guo, and G. Zhu, 
``Hybrid active-passive RIS transmitter enabled energy-efficient multi-user communications,'' 
\emph{IEEE Trans. Wireless Commun.}, 
vol. 23, no. 9, pp. 10653$-$10666, Sep. 2024.


\bibitem{ref_TRIS_app_5}
Z. Liu \emph{et al.}, 
``Beamforming design and multi-user scheduling in transmissive RIS enabled distributed cooperative ISAC networks with RSMA,'' 
\emph{IEEE Trans. Commun.}, 
vol. 73, no. 12, pp. 15247$-$15263, Dec. 2025.



\bibitem{ref_TRIS_app_6}
Z. Liu, W. Chen, Z. Li, and Q. Wu, 
``Joint spatial registration and resource allocation for transmissive RIS enabled cooperative ISCC networks,'' 
\emph{IEEE Trans. Commun.}, 
vol. 74, pp. 7494$-$7509, 2026.

\bibitem{ref_TRIS_app_7}
Y. Guo, W. Chen, Q. Wu, Y. Zhu, Y. Liu, Z. Li, and Y. Wang,
``Max-min rate optimization for multigroup multicast MISO systems via novel transmissive RIS transceiver,'' 
Jul. 2025.
[Online]. 
Available: https://arxiv.org/abs/2507.18733


\bibitem{ref_TRIS_app_8}
Y. Guo \emph{et al.}, 
``Fair rate maximization for multi-user multi-cell MISO communication systems via novel transmissive RIS transceiver,'' 
\emph{IEEE Trans. Veh. Technol.}, 
early access,
January 12, 2026,
doi: 10.1109/TVT.2026.3651497.



\bibitem{ref_TRIS_app_9}
A. Umra, A. M. Ahmed, S. Roth, and A. Sezgin, 
``Hardware-efficient cognitive radar: Multi-target detection with RL-driven transmissive RIS,''
in \emph{Proc. IEEE Int. Conf. Acoust., Speech Signal Process. (ICASSP)}, Barcelona, Spain, May 2026, pp. 20621$-$20625.

\bibitem{ref_TRIS_app_10}
M. Boloori, C. Li, and A. Sezgin,
``Optimizing movable antenna position and transmissive RIS phase for efficient base station design,''
Nov. 2025.
[Online]. 
Available: https://arxiv.org/abs/2511.01575


\bibitem{ref_channel_error_1}
Y. Liu, J. Li, and H. Wang, 
``Robust linear beamforming in wireless sensor networks,''
\emph{IEEE Trans. Commun.}, 
vol. 67, no. 6, pp. 4450$-$4463, Jun. 2019.


\bibitem{ref_FP}
K. Shen and W. Yu, 
``Fractional programming for communication systems-part I: power control and beamforming,'' 
\emph{IEEE Trans. Signal Process.}, 
vol. 66, no. 10, pp. 2616$-$2630, May 2018.

\bibitem{ref_FP_1}
K. Shen and W. Yu, 
``Fractional programming for communication systems-part II: Uplink scheduling via matching,'' 
\emph{IEEE Trans. Signal Process.}, 
vol. 66, no. 10, pp. 2631$-$2644, May 2018.


\bibitem{ref_MM}
Y. Sun, P. Babu, and D. P. Palomar,
``Majorization-minimization algorithms in signal processing, communications, and machine learning,''
\emph{IEEE Trans. Signal Process.},
vol. 65, no. 3, pp. 794$-$816, Feb. 2017.


\bibitem{ref_S_Procedure}
Y. C. Eldar and N. Merhav, 
``A competitive minimax approach to robust estimation of random parameters,'' 
\emph{IEEE Trans. Signal Process.}, 
vol. 52, no. 7, pp. 1931$-$1946, Jul. 2004.


\bibitem{ref_joint_signal}
X. Liu \emph{et al.}, 
``Joint transmit beamforming for multiuser MIMO communications and MIMO radar,'' 
\emph{IEEE Trans. Signal Process.}, 
vol. 68, pp. 3929$-$3944, 2020.



\bibitem{ref_channel_estimation_1}
Z. Li \emph{et al.}, 
``Transmissive RIS transceiver enabled multistream communication systems: Design, optimization, and analysis,'' 
\emph{IEEE Internet Things J.}, 
vol. 12, no. 5, pp. 5985$-$6000,  Mar. 2025.


\bibitem{ref_channel_estimation_2}
G. Zhou, C. Pan, H. Ren, P. Popovski, and A. L. Swindlehurst, 
``Channel estimation for RIS-aided multiuser millimeter-wave systems,'' 
\emph{IEEE Trans. Signal Process.}, 
vol. 70, pp. 1478$-$1492, Mar. 2022.

\bibitem{ref_beampattern}
M. Hua, Q. Wu, C. He, S. Ma, and W. Chen, 
``Joint active and passive beamforming design for IRS-aided radar-communication,'' 
\emph{IEEE Trans. Wireless Commun.}, 
vol. 22, no. 4, pp. 2278$-$2294, Apr. 2023.


\bibitem{ref_beampattern_1}
Z. Zheng, Q. Wu, Y. Zhu, W. Chen, Y. Gao, and H. Wang, 
``Wireless sensing with movable intelligent surface,'' 
\emph{IEEE J. Sel. Topics Signal Process.}, 
early access,
April 06, 2026,
doi: 10.1109/JSTSP.2026.3681476.


\bibitem{ref_P_cir}
E. Bj\"{o}rnson, J. Hoydis, and L. Sanguinetti, 
``Massive MIMO networks: Spectral, energy, and hardware efficiency,'' 
\emph{Foundat. Trends Signal Process.}, vol. 11, no. 3$-$4, pp. 154$-$655, 2017.

\bibitem{ref_channel_error_model}
G. Zhou, C. Pan, H. Ren, K. Wang, and A. Nallanathan, 
``A framework of robust transmission design for IRS-aided MISO communications with imperfect cascaded channels,'' 
\emph{IEEE Trans. Signal Process.}, 
vol. 68, pp. 5092$-$5106, 2020.

\bibitem{ref_channel_error_model_1}
D. Xu, X. Yu, D. W. K. Ng, A. Schmeink, and R. Schober, 
``Robust and secure resource allocation for ISAC systems: A novel optimization framework for variable-length snapshots,'' 
\emph{IEEE Trans. Commun.}, 
vol. 70, no. 12, pp. 8196$-$8214, Dec. 2022.


\bibitem{ref_channel_error_model_2}
{T. -X. Zheng \emph{et al.}, 
``Reconfigurable intelligent surface-aided secure integrated radar and communication systems,'' 
\emph{IEEE Trans. Wireless Commun.}, 
vol. 24, no. 3, pp. 1934$-$1948, Mar. 2025.}


\bibitem{ref_BCA}
D. P. Bertsekas, 
``Nonlinear programming,'' 
\emph{Journal of the Operational Research Society}, 
vol. 48, no. 3, pp. 334$-$334, 1997.

\bibitem{ref_EE_joint_convex}
Y. Zhu, Y. Liu, M. Li, Q. Wu, and Q. Shi, 
``A flexible design for active reconfigurable intelligent surface-A sub-array architecture,'' 
\emph{IEEE Trans. Veh. Technol.}, 
vol. 72, no. 10, pp. 12884$-$12899, Oct. 2023.


\bibitem{ref_CVX}
M. Grant and S. Boyd,
\emph{CVX: Matlab software for disciplined convex programming}, 
version 2.1, http://cvxr.com/cvx, Mar. 2014.

\bibitem{ref_complexity}
A. Ben-Tal and A. Nemirovski, 
``Lectures on Modern Convex Optimization, Analysis, Algorithms, and Engineering Applications.'' 
\emph{Society for Industrial and Applied Mathematics (SIAM)}, 2001.

\bibitem{ref_sign}
I. R. Petersen, 
``A stabilization algorithm for a class of uncertain linear systems,'' 
\emph{Syst. Control Lett.}, 
vol. 17, no. 2, pp. 351$-$357, 1987.


\bibitem{ref_Convex Optimization}
S. Boyd and L. Vandenberghe,
\emph{Convex Optimization.}
New York: Cambridge University Press, 2004.



\end{thebibliography}
\end{document}